\documentclass[prb,showpacs,floatfix,superscriptaddress,twocolumn,aps,10]{revtex4-2}
\usepackage{xcolor}
\usepackage{tikz}
\usepackage{polyhedra}
\usepackage{extarrows}
\usepackage{graphicx}
\usepackage{caption}
\usepackage[inline]{enumitem}
\usepackage{amsmath}
\usepackage{lmodern,amsmath,amssymb,mathtools,booktabs}
\usepackage{parskip}
\usepackage{bm}
\usepackage[hidelinks]{hyperref}
\newcommand{\av}[1]{\left\langle #1\right\rangle}
\newcommand{\E}{\mathbb E}
\newcommand{\Prob}{\mathbb P}
\newcommand{\BZ}{\mathrm{BZ}}
\newcommand{\Id}{\mathbb I}
\usepackage{comment}
\numberwithin{equation}{section} 
\usepackage[version=4]{mhchem} 
\usepackage{subcaption}

\usepackage{graphicx}
\usepackage{dcolumn}
\usepackage{bm}

\begin{document}

\preprint{APS/123-QED}

\title{Spin Structure Factor of the classical Ising Model on the Centered Pyrochlore Lattice }

    \author{Lechen Xu}
	\email[]{Xu.Lechen@lmu.de}
	\affiliation{Department of Physics and Arnold Sommerfeld Center for Theoretical Physics (ASC), Ludwig Maximilian University of Munich, 80333 Munich, Germany}
  
    \author{Tomas Krei\ss el}
	\email[]{Tomas.Kreissel@lmu.de}
	\affiliation{Department of Physics and Arnold Sommerfeld Center for Theoretical Physics (ASC), Ludwig Maximilian University of Munich, 80333 Munich, Germany}

    \author{Naïmo Davier}
	\email[]{naimo.davier@u-bordeaux.fr}
    \affiliation{CNRS, Universit\'e de Bordeaux, LOMA, UMR 5798, 33400 Talence, France}

    \author{Ludovic D.C. Jaubert}
	\email[]{ludovic.jaubert@u-bordeaux.fr}
    \affiliation{CNRS, Universit\'e de Bordeaux, LOMA, UMR 5798, 33400 Talence, France}
    
    \author{Lode Pollet}
	\email[]{Lode.Pollet@lmu.de}
	\affiliation{Department of Physics and Arnold Sommerfeld Center for Theoretical Physics (ASC), Ludwig Maximilian University of Munich, 80333 Munich, Germany}
\affiliation{Munich Center for Quantum Science and Technology (MCQST), 80799 Munich, Germany}
   
\date{\today}

\begin{abstract}
We present the spin structure factor of the Ising model on the centered pyrochlore lattice. Generalizing Henley's method~\cite{PhysRevB.71.014424} to the case of inequivalent center and vertex spins, we obtain spin correlation functions that are in excellent agreement with Monte Carlo results and can capture sharp and broadened pinch points, also at finite temperature. Our approach can be extended to other centered lattice structures in classical frustrated magnetism.
\end{abstract}

\maketitle


\section{Introduction}
\label{chapter:intro}

Geometrical frustration can prevent magnetic ordering giving rise to extensively degenerate ground-states governed by local constraints. A prominent example is spin ice, realized in several rare-earth pyrochlore oxides, and implemented in its simplest form at the  classical level by taking Ising spins on the pyrochlore lattice \cite{harris1997,udagawa2021}. Its two-in--two-out constraint admits a description in terms of an emergent Coulomb phase with a divergence-free field showing algebraic correlations and featuring characteristic pinch points in the structure factor. Violations of this constraint carry emergent gauge charge in the form of magnetic monopoles \cite{castelnovo2007,https://doi.org/10.1002/cphc.200900873}. 

\begin{figure}[t]
    \centering
    \includegraphics[width=0.8\linewidth]{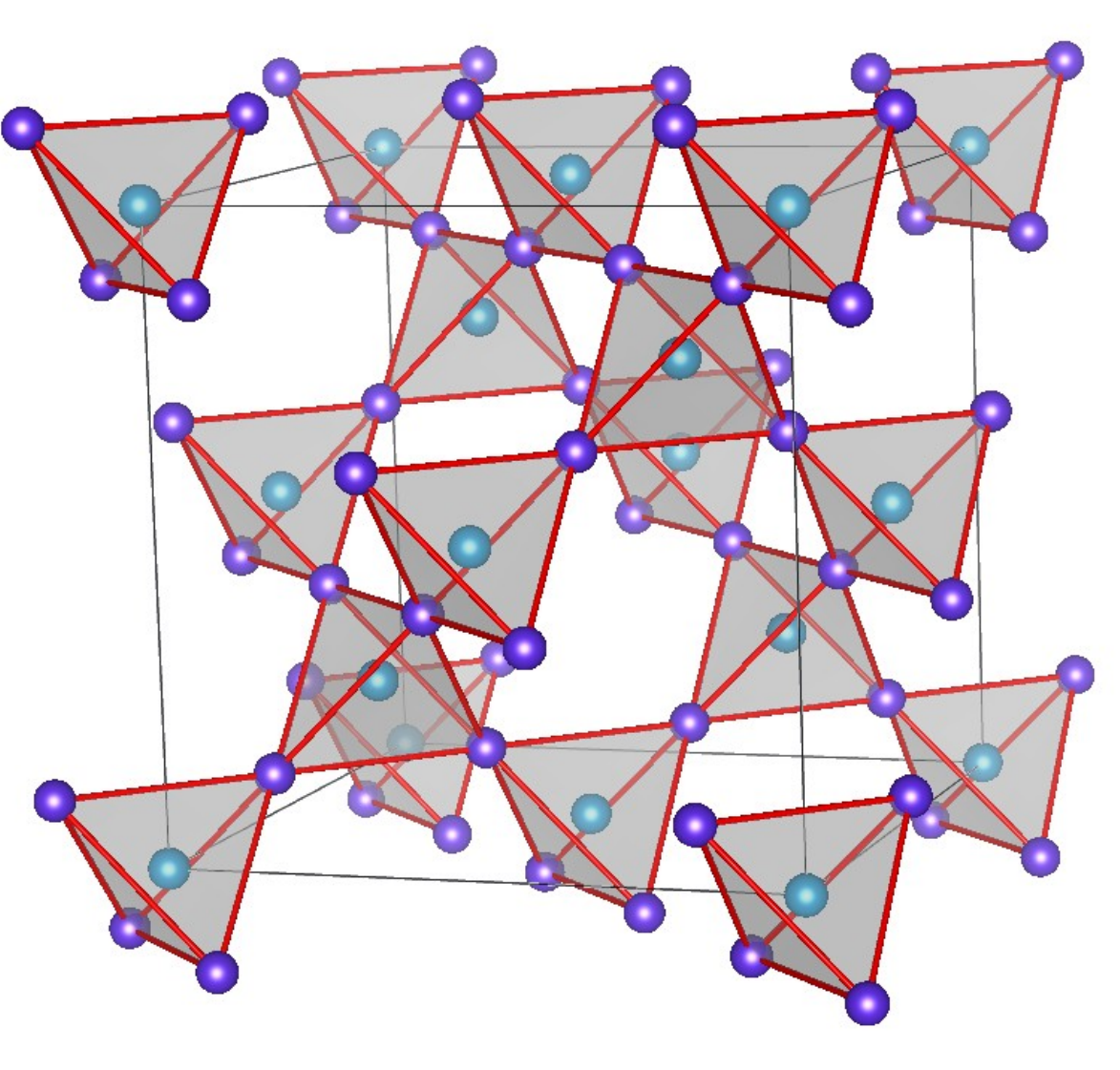}
    \caption{The conventional unit cell of the centered pyrochlore lattice is shown as a grey cube, containing 4 primitive unit cells and $6 \times 4 = 24$ sublattices. The purple sites are the vertices of the tetrahedra, and the cyan sites are the centers. Every cyan center is itself a vertex of the diamond lattice, and the purple vertices are located in the middle of the bonds of the diamond lattice. The diamond lattice bond connects the centers of one a-type and one b-type tetrahedron.  }
    \label{fig:CPy}
\end{figure}

The centered pyrochlore lattice extends this framework by placing an additional spin at the center of every tetrahedron, increasing the number of sublattices from four to six. This geometry occurs in the metal-organic framework Mn(ta)$_2$, where (ta) denotes the 1,2,3-triazolate ligand \cite{gandara2012,NutakkiScipost2023}. As opposed to traditional spin-ice physics, competition between center--vertex and vertex--vertex exchange interactions allows charged tetrahedral configurations to enter the ground-state manifold. For the Ising model, the ground-state phase diagram contains a monopole fluid \cite{NutakkiScipost2023}, in which the vertex spins on the pyrochlore lattice sites obey the $3-1$ rule, and the center spin aligns with the minority spin.

This raises a concrete question: how do the local constraints and the additional center-spin fluctuations determine the spin structure factor? In this work, we provide a unified framework, valid for spin-ice as well as for the monopole fluid, to  calculate vertex--vertex, center--vertex, and center--center correlations at zero and finite temperature using exact conditional identities together with a self-consistent Gaussian approximation building on Henley's projector method \cite{PhysRevB.71.014424}. In fact, the spin correlation function in the monopole fluid is known exactly. Comparisons with Monte Carlo simulations test the quantitative accuracy of this description near $\gamma=1$. For smaller coupling ratios, we extend the construction to a three-level description which improves agreement with Monte Carlo results over the two-level approximation to higher temperatures.

\begin{figure}[t]
    \centering
        \includegraphics[width=\linewidth]{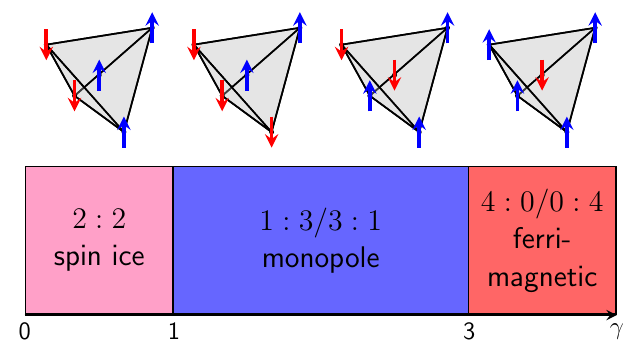}
        \caption{Phase diagram of the classical Ising model on CPy, with examples of the tetrahedra configuration above each phase region. The names of the phases are the same as in Refs.~\cite{NutakkiScipost2023,Nutakki2025}. When $\gamma < 1$, the system is in the spin ice state ($2{:}2$), in which every tetrahedron has 2 up vertex spins, 2 down vertex spins, and a randomly oriented center spin. When $1 < \gamma < 3$, the vertex spins are in either the 3-up-1-down/1-up-3-down configuration, and the center spin takes the same value as the minority vertex spin. We refer to this as the monopole fluid ($3{:}1{/}1{:}3, L$), where $L$ stands for ``low energy'', indicating that the center spin orientation minimizes the energy, i.e. it points along the minority spin. On the opposite, if the center spin was reversed, it would correspond to the high-energy state, labeled $(3:1/1:3,H)$. When $\gamma >3$, the system forms a ferrimagnetic state where all the vertex spins are aligned, and the center spins are oriented in the opposite direction ($4{:}0{/}0{:}4, L$). At the two special points, $ \gamma= 1$ and $3$, the phases on both sides of these points coexist. }
     \label{subfig:Ising_model_zero_phase}
 \end{figure}


\section{Model}
\label{chapter:model}

\subsection{\label{sec:lattice and model}Centred Pyrochlore lattice and the Model}

The centered pyrochlore (CPy) lattice structure, shown in Fig.~\ref{fig:CPy} has recently received some attention in the literature, {\it e.g.}, for the classical Heisenberg model~\cite{NutakkiScipost2023}, the quantum spin-1/2 XXZ model~\cite{Nutakki2025}, and the classical Heisenberg model with dipolar interactions~\cite{NutakkiPRR2023}. Here, we consider the $J_1-J_2$ antiferromagnetic classical Ising model
\begin{align}
    H_{\rm Ising} &= J_1\sum_{\langle i,j \rangle} c_i v_j + J_2\sum_{\ll i,j \gg} v_i v_j\\
    &=\sum_{t}\frac{J_2}{2} \left( \gamma c_t  + \sum_{j \in t} v_j \right)^2 + {\rm const} \label{eq:II.A.Hamiltonian}\\
    &= \frac{J_2}{2}\sum_t \left( \gamma c_t + m_t \right)^2 + {\rm const},  
    \label{eq:II.A.Hamiltonian_2}
\end{align}
where $J_1$ is the nearest-neighbor interaction strength between a center and a vertex spin and $J_2$ is the next-nearest-neighbor interaction strength between two vertices. Up and down tetrahedra are denoted as $t=\{a,b\}$, center spins as $c_{t}$, vertex spins as $v_{j=\{0,1,2,3\}}$.  The spins take Ising values,  $c,v \in \{-1, +1\}$.  The magnetization over the vertex spins per tetrahedron $t$ is $m_t=\sum_{i\in t} v_i$. We take $\gamma = \frac{J_1}{J_2}$ to parameterize the model, and set the unit $J_2 = J = 1$.

\subsection{\label{sec:mapping}Ground states, their energies, and degeneracy}

The ground state phase diagram has been discussed in Ref.~\cite{NutakkiScipost2023} and is repeated in Fig.~\ref{subfig:Ising_model_zero_phase}. The terminology of the three phases is discussed in the figure caption.

The relevant local levels are
\begin{center}
\small
\renewcommand{\arraystretch}{1.3}
\begin{tabular*}{\columnwidth}{@{\extracolsep{\fill}}clcc}
\toprule
\normalsize Level & \normalsize Condition & \normalsize Energy & \normalsize Degeneracy\\
\midrule
$2{:}2$          & $m_t=0$                 & $\frac{J\gamma^2}{2}$     & $12$\\
$3{:}1/1{:}3,L$  & $|m_t|=2,\ c_t m_t=-2$  & $\frac{J(2-\gamma)^2}{2}$ & $8$\\
$3{:}1/1{:}3,H$  & $|m_t|=2,\ c_t m_t=+2$   & $\frac{J(2+\gamma)^2}{2}$ & $8$\\
$4{:}0/0{:}4,L$  & $|m_t|=4,\ c_t m_t=-4$  & $\frac{J(4-\gamma)^2}{2}$ & $2$\\
$4{:}0/0{:}4,H$  & $|m_t|=4,\ c_t m_t=+4$  & $\frac{J(4+\gamma)^2}{2}$ & $2$\\
\bottomrule
\end{tabular*}
\label{table:GS}
\end{center}

\begin{figure}
    \centering
    \includegraphics[width=1.0\linewidth]{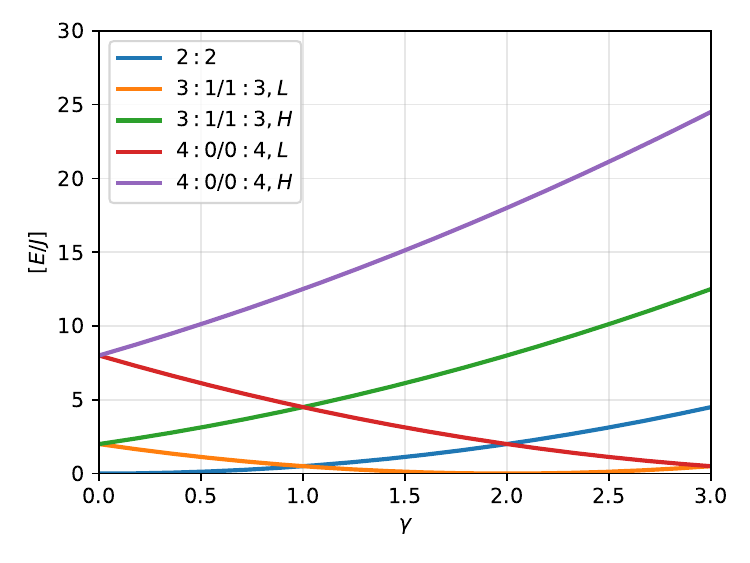}
    \caption{Energy levels of the possible spin configurations on an independent tetrahedron, visualizing the lowest excitations in spin-ice and monopole fluid, respectively. Ferro- and ferrimagnetic correlations ($4{:}0/0{:}4$) cannot be dealt with in our analytical treatment; all other spin-liquid-like configurations can be taken into account. }
    \label{fig:energy_levels}
\end{figure}

This gives rise to the energy spectrum shown in Fig.~\ref{fig:energy_levels}.


The main quantity of interest of this paper is the static spin structure factor, which is the Fourier transform of the spin correlation function, and which we split into a vertex--vertex, center--vertex, and center--center part,
\begin{eqnarray}
S(\mathbf{k}) & = & \sum_{i,j} S_{v_iv_j}(\mathbf{k}) + \sum_{\alpha j} S_{c_\alpha v_j}(\mathbf{k}) + \sum_{\alpha \beta} S_{c_\alpha c_\beta}(\mathbf{k})  \nonumber, \\
S_{v_iv_j}(\mathbf{k}) & = & \langle v_i(-\mathbf{k}) v_j(\mathbf{k}) \nonumber \rangle,\\ 
S_{c_\alpha v_j}(\mathbf{k}) & = & \langle c_\alpha(-\mathbf{k}) v_j(\mathbf{k}) \nonumber \rangle,\\
S_{c_\alpha c_\beta}(\mathbf{k}) & = & \langle c_{\alpha}(-\mathbf{k}) c_{\beta}(\mathbf{k})\rangle,
\end{eqnarray}
where the sums run over $i,j=0, \ldots 3$ and $\alpha,\beta =a,b$.
The main results of this paper are the closed form expressions for the spin correlation functions which are seen to be in excellent agreement with Monte Carlo simulations:
\begin{enumerate}
\item For $0 < \gamma < 1$ we can take the 3 lowest energy levels into account and be accurate up to temperatures $ T \lesssim 3J$.
\item For $\gamma = 1$ we take the two degenerate ground state levels into account. In that case, temperature effects are absent.
\item For $1 < \gamma < 2$ we can take the two lowest energy levels into account, which also allows us to accurately calculate the correlation function at low temperatures and have an exact expression in the ground state.
\item For $2 < \gamma < 3$ we can provide the exact ground state results.
\end{enumerate}
We start with the analytical expressions for the correlation function at zero temperature.


\section{Spin Correlation Function at  Zero Temperature}
\subsection{Spin ice: $0<\gamma<1$}
In this case, every tetrahedron has a zero vertex magnetization, $m_t=0$, from the 2-in 2-out spin ice rule. The values of the center spins can be chosen freely, so 
\begin{equation}
\av{c_tc_{t'}}=\delta_{tt'},\qquad \av{c_tv_i}=0.
\end{equation}
For the vertex correlations, we will use Henley's Gaussian approximation~\cite{PhysRevB.71.014424}
\begin{equation}
P_G[v]\propto
\exp\left[-\frac{1}{2\sigma_v^2}\sum_i v_i^2\right]
\prod_t\delta\left(\sum_{i\in t}v_i\right),
\quad \av{v_i^2}_G=1.
\end{equation}
The decoupled discrete center distribution supplies an additional factor
$2^{-N_t}$ in the joint measure. The vertex-vertex spin correlator is proportional to the projector operator, projecting the fluctuating spins onto the physical, Gauss' law-abiding space. Henley's method is known to provide a very good approximation to the spin correlation function in the spin ice phase. As the ground state correlator in the spin ice phase is identical to the one derived by Henley, we will not go into details  here. Note, however, that the zero-temperature correlator will follow as a special case in the limit $T \to 0$ and $\gamma < 1$ of an extension of Henley's formalism presented below.

\subsection{Monopole fluid: $1<\gamma<3$}
Remarkably, the spin correlation function can be written down analytically in the ground state of the monopole fluid.
The allowed tetrahedra are in the  $3{:}1/1{:}3,L$ configurations, in which the product over the vertex spins has negative parity and the center spin is fixed by the vertex magnetization,
\begin{equation}
\prod_{i\in t}v_i=-1,\qquad c_t=-\frac12\sum_{i\in t}v_i.
\label{eq:parity}
\end{equation}
Consider a closed loop on the diamond lattice and flip the vertex spins along the loop. 
Such a loop flips two spins per visited tetrahedron, thus preserving its parity. For distinct sites $i,j$, choose a loop
containing $i$ but not $j$. The transformation changes $v_iv_j$ to
$-v_iv_j$ and preserves the ensemble weight. Hence,
\begin{equation}
\av{v_i}=0,\qquad \av{v_iv_j}=\delta_{ij}.
\end{equation}
Using Eq.~\eqref{eq:parity}, it follows that
\begin{align}
\av{c_tv_i}&=
\begin{cases}-1/2,&i\in t,\\0,&i\notin t,\end{cases}\\
\av{c_tc_{t'}}&=
\begin{cases}
1,&t=t',\\
1/4,&t,t'\text{ nearest neighbors},\\
0,&\text{otherwise}.
\end{cases}
\end{align}
This proves that the spin correlations in the monopole ground state are known exactly.


\section{Gaussian construction for the reduced variables}
\label{sec:Gauss}

As listed in the above table, any tetrahedron can be one of five possible states. In this paper, we will not consider 4:0/0:4,($L$ or $H$) states, as they do not fit within our formalism. It is this condition that determines the region of validity of our calculations, since 4:0/0:4,$L$ become the first excited states for $\gamma>2$ and the ground states for $\gamma>3$. Nevertheless, our theory is valid at low temperatures for the most interesting parameter region $0\leq \gamma < 2$, and at zero temperature up to $\gamma < 3$, as vindicated by Monte Carlo simulations in Sec.\ref{chapter:MC}. In the following sections, we will first consider the neutral 2:2 and monopole 3:1/1:3,$L$ states (two levels) and then include 3:1/1:3,$H$ excitations (three levels).

\subsubsection{\label{subsec:two-level}Two-level ensemble for $0<\gamma<2$.}

The center spin behaves differently in the spin-ice and in the monopole fluid state.
In the former, it is free to fluctuate in the ground state; in the latter, its value is deterministic.
A formalism capable of dealing with both states requires that we split the center spin into a deterministic term $s_t$ and a residual term $r_t$, 
\begin{equation}
s_t=-\frac12m_t\in\{0,\pm1\},\qquad c_t=s_t+r_t.
\label{eq:stdef}
\end{equation}
For a neutral 2:2 tetrahedron, $s_t=0$ and $r_t=c_t$; for a 3:1/1:3,$L$ monopole,
$s_t=c_t$ and $r_t=0$. In particular, $s_tr_t=0$ in every configuration. 
Mathematically speaking, the dependence of the center spin on the magnetization of the vertex spins and the type of spin liquid configuration can be expressed as a conditional probability. 
We note that, conditioned on the vertex spins, the residual terms on neutral tetrahedra ($2{:}2$)  are independent and free, whereas for monopoles, there are no residual terms. We can hence combine these statements into an expectation value subject to a vertex configuration condition as
\begin{equation}
\E[r_t\mid\{v\}]=0,\qquad
\av{s_tr_{t'}}=0,\qquad \av{v_ir_t}=0.
\end{equation}
Let $n$ be the monopole fraction in the statistical mixture. Then
\begin{equation}
\av{s_t^2}=n,\qquad
\av{r_tr_{t'}}=(1-n)\delta_{tt'}.
\label{eq:norm_s_r}
\end{equation}
It follows that
\begin{align}
\av{c_tv_i}&=\av{s_tv_i},\\
\av{c_tc_{t'}}&=\av{s_ts_{t'}}+(1-n)\delta_{tt'}.
\label{eq:two-reconstruction}
\end{align}
The normalization conditions
\begin{equation}
\av{v_i^2}=1,\qquad \av{s_t^2}=n \, ,
\label{eq:norm_v_s}
\end{equation}
ensure $\av{c_t^2}=n+(1-n)=1$.
The calculation of the $\av{s_tv_i}$ and $\av{v_iv_j}$ correlations is the goal of the Gaussian construction exposed in the rest of the section.

Following Henley~\cite{PhysRevB.71.014424}, we promote $(v,s)$ to continuous variables and use Gaussian fluctuations which are independent per tetrahedron but subject to the Gauss' law constraint per tetrahedron, 
\begin{eqnarray}
P_G[v,s] & \propto & \exp\left[-\frac{1}{2A}\sum_i v_i^2-\frac{1}{2B}\sum_t s_t^2\right] 
\prod_t\delta G_t \nonumber, \\
G_t & = & \left(\sum_{i\in t}v_i+2s_t\right).
\label{eq:gaussian}
\end{eqnarray}
The Gauss' law constraint $G_t=0$ is always valid in the two-level system, by definition of Eq.~(\ref{eq:stdef}). Note that we wrote the center spin degrees of freedom immediately in the  $s$ variables, as the freely fluctuating $r$ variables only contribute to the normalization. The constraint is an expression valid for both types of spin liquids, and, importantly, it is linear.
Since the constraint affects the Gaussian fluctuations, $A,B$ are not the physical variances of the real spin fluctuations, but a useful intermediary in our formalism.
With the rescaling
\begin{equation}
s_t=x\widetilde s_t,\qquad x=\sqrt{B/A},
\label{eq:rescaling}
\end{equation}
the statistical weight becomes isotropic with width $A$, and the constraint is
\begin{equation}
\sum_{i\in t}v_i+2x \widetilde s_t=0.
\end{equation}
At this point we go over to momentum space.
Let $\Pi(\mathbf k; x)$ be the orthogonal projector onto the allowed
Fourier subspace of Gauss' law satisfying tetrahedra, in the labeling $(\widetilde s_a,\widetilde s_b,v_0,v_1,v_2,v_3)$.
The momentum space representation of the constraint considered over the full Brillouin zone is
\begin{equation}
    G^{\pm}(\mathbf{k}) = \sum_{i=0}^3e^{\pm i \mathbf{k}\cdot\bm{\delta}_i} v_i (\mathbf{k}) + 2 x \tilde{s}_{a,b}(\mathbf{k}),
   \label{eq:constraint_k}
\end{equation}
where $\pm$ corresponds to the $a(b)$ sublattice of diamond and the position vectors $\bm{\delta}_i$ are defined in Appendix \ref{Appendix:Lattice Definition}. Thus, we have six degrees of freedom with two constraints, implying that $\Pi(\mathbf k; 2x)$ has rank four for non-zero $x$. 
The constraints are stacked into the $6\times2$ matrix
\begin{equation}
E_c(\mathbf k;2x)=(G^{+}(\mathbf{k}),G^{-}(\mathbf{k})).
\end{equation}
The projector operator (a full derivation is provided in Appendix~\ref{Appendix:Projector operator}) expressed in terms of the $E_c$ is then
\begin{equation}
\Pi=\Id_6-E_c(E_c^{\dagger}E_c)^{-1}E_c^{\dagger}.
\end{equation}


The approximate spin correlation function can be obtained from $\Pi$ when also taking the normalization conditions (Eq.~\ref{eq:norm_s_r} and Eq.~\ref{eq:norm_v_s}) into account. 
To this end we introduce the normalized Brillouin-zone averages
\begin{align}
P_s(2x)&=\int_{\BZ}\frac{d^3k}{V_{\BZ}}\,
\Pi_{aa}(\mathbf k; 2x),\\
P_v(2x)&=\int_{\BZ}\frac{d^3k}{V_{\BZ}}\,
\Pi_{00}(\mathbf k; 2x ).
\end{align}
Sublattice symmetry makes the two center diagonal averages equal, and
likewise the four vertex diagonal averages. 
Per Eq.~\ref{eq:norm_s_r} and Eq.~\ref{eq:norm_v_s}, 
the residual terms $r$ of the center spins contribute a constant factor $(1-n)$ since $r$ does not enter Eq.~\ref{eq:gaussian}; the deterministic terms $s$ of the center spins enter through the projector $\Pi$ and the variance $B$ normalized selfconsistently to $n$, and the vertex spins enter through the projector $\Pi$ and the variance $A$ normalized selfconsistently to $1$:
\begin{equation}
BP_s( 2x )=n,\qquad AP_v( 2x)=1,
\end{equation}
so that (see Eq.~\ref{eq:rescaling})
\begin{equation}
x^2=n\,\frac{P_v(2x)}{P_s(2x)}.
\end{equation}
For nonzero $x$, the projector has rank four, and therefore
\begin{equation}
2P_s+4P_v=4,\qquad
A=1+\frac{n}{2x^2},\qquad B=x^2+\frac n2.
\end{equation}

With the notation
\begin{equation}
D=\operatorname{diag}(\sqrt B,\sqrt B,\sqrt A,\sqrt A,\sqrt A,\sqrt A),
\, \, 
R=\begin{pmatrix}\Id_2&0\\0&0_4\end{pmatrix},
\end{equation}
the Gaussian reduced covariance is $D\Pi D$. 
The analytical approximation to the spin correlation function is then found as
\begin{equation}
C(\mathbf k)=D\Pi(\mathbf k;2x)D+(1-n)R.
\label{eq:matrix-two}
\end{equation}
We note again that the center--center correlation has the constant background $(1-n)\Id_2$, absent for the center--vertex and vertex--vertex contribution, because of the unconstrained fluctuations of center spins within the two-in--two-out portion of tetrahedra.

\begin{figure}
    \centering
    \includegraphics[width=1.0\linewidth]{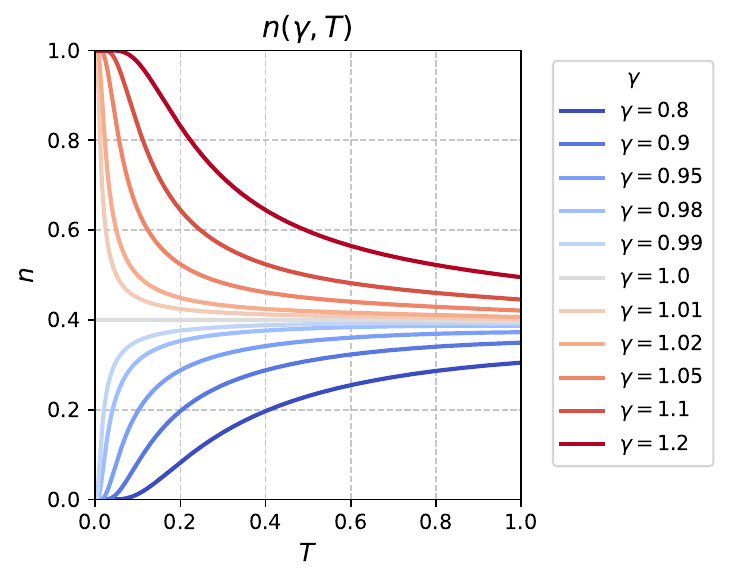}
    \caption{Plot of the density of $3:1/1:3, L$ tetrahedra (see Appendix~\ref{App:Stat_Mech}).  In this figure, we take $0.8 < \gamma < 1.2$, and $T \leq 1$. 
    }
    \label{fig:Figs/CPy_n_gamma_T.pdf}
\end{figure}

The previous discussion is valid for any monopole fraction $n$; in particular it remains valid at finite temperature where $n$ follows from single-tetrahedron equilibrium mechanics: Within a two-level ensemble per tetrahedron ($2{:}2$ and $3{:}1/1{:}3,L$ tetrahedra), temperature and $\gamma$ enter through $n(\gamma,T)$ as shown in Fig.\ref{fig:Figs/CPy_n_gamma_T.pdf}. The excitation energies, partition function, and level population are calculated in Appendix~\ref{App:Stat_Mech}.

Specific for the degenerate $\gamma=1$ point, local degeneracy counting (see Eq.~\ref{eq:Boltzmann1} and Eq.~\ref{eq:Boltzmann0}) gives $n=8/(12+8)=2/5$. The $\gamma=1$ point is therefore just a limiting case of our more general framework. Note that at $\gamma=1$, the truncated ensemble has no temperature dependence since all 20 states are degenerate.

\subsubsection{\label{subsec:Finite temperature within the three-level ensemble}Three-level ensemble for small $0<\gamma<1$}

Remarkably, it is possible to also include the $3{:}1/1{:}3,H$ excitations in the Gaussian approach and form a three-level ensemble. 
This is only meaningful on top of the spin-ice ground state $(2{:}2)$. The equilibrium statistical mechanics description is again provided in Appendix~\ref{App:Stat_Mech}.

As before, we write $ s_t=-m_t/2\in\{0,\pm1\}$.  On charged tetrahedra, $c_t=s_t$ in the $3{:}1/1{:}3,L$ configurations and
$c_t=-s_t$ in the $3{:}1/1{:}3,H$ configurations.


The energy between center and vertex spins can be written in compact form as
$J\gamma c_tm_t=-2J\gamma c_ts_t$. 
This allows one to write the probability that the center spins takes a particular value conditioned on the vertex configuration for a charged tetrahedron as follows:
For $s_t=\pm1$,
\begin{align}
\Prob(c_t= \pm 1 \mid\{v\} \quad {\rm s.t} \quad s_t = \pm 1 )&=
\frac{e^{2\beta J\gamma}}{2\cosh(2\beta J\gamma)},\\
\Prob(c_t= \mp 1 \mid\{v\} \quad {\rm s.t} \quad s_t = \pm 1 )&=
\frac{e^{-2\beta J\gamma}}{2\cosh(2\beta J\gamma)}.
\end{align}
For the ground state we have additionally $\Prob(c_t= \pm 1 \mid\{v\} \, \, {\rm s.t.} \,\,  s_t = 0 ) = \frac12$.
Consequently,
\begin{equation}
\E[c_t\mid\{v\} \quad {\rm s.t} \quad s_t = \pm 1  ]= \pm a,  \qquad a=\tanh(2\beta J\gamma) .
\end{equation}
or, directly, $\E[c_t\mid\{v\}   ]=  a s_t$, valid for excited and ground states. 
Thanks to this linear relationship, we can again decompose the stochastic variable for the center spin into a deterministic part (which is fixed by the vertex spins) and a fluctuating residual part,
\begin{equation}
c_t=a s_t+r_t,\qquad {\rm with} \quad \E[r_t\mid\{v\}]=0.
\end{equation}
The expectation value of the cross term, conditioned on the vertex spins, also vanishes,
\begin{equation}
\E[r_ts_t\mid\{v\}]
=s_t\E[c_t\mid\{v\}]-a s_t^2=0.
\end{equation}
More generally, $\av{r_tF(\{v\})}=0$ for any function of the vertices.

Expanding the conditional second moment gives
\begin{align}
\E[r_t^2\mid\{v\}]
&=\E[c_t^2-2a s_tc_t+a^2s_t^2\mid\{v\}]\\
&=1-2a s_t(a s_t)+a^2s_t^2\\
&=1-a^2s_t^2.
\end{align}
The centers fluctuate independently. With
$n=n_L+n_H=\av{s_t^2}$, this yields
\begin{equation}
\av{r_tr_{t'}}=(1-a^2n)\delta_{tt'}.
\end{equation}
Therefore,
\begin{align}
\av{c_tv_i}=a\av{s_tv_i},\\
\av{c_tc_{t'}}=
a^2\av{s_ts_{t'}}+(1-a^2n)\delta_{tt'}.
\end{align}
As a check, we see that the physical center is correctly normalized,
\begin{equation}
\av{c_t^2}=a^2n+(1-a^2n)=1.
\end{equation}

The Gaussian projector construction is otherwise unchanged; the equations to be satisfied are
$ m_t+2s_t=0, \av{v_i^2}_G=1,$ and $\av{s_t^2}_G=n$.
The expression for the spin correlation function is then found by only taking the parameter $a$ and the new values of $n$ into account compared to the previous expression for 2-level systems,
\begin{eqnarray}
D_a & = & \operatorname{diag}(a\sqrt B,a\sqrt B,\sqrt A,\sqrt A,\sqrt A,\sqrt A), \nonumber \\
C(\mathbf k) & = & D_a\Pi(\mathbf k;2x)D_a+(1-a^2n)R.
\end{eqnarray}
It can be seen that, as $\gamma \to 0$ at fixed $T$, $a \to 0$, in which case the center spins fluctuate freely.


\section{Monte Carlo Simulation}
\label{chapter:MC}

\begin{figure}[t]
    \centering
    \begin{subfigure}[b]{1.0\linewidth} 
        \centering
        \includegraphics[width=\linewidth]{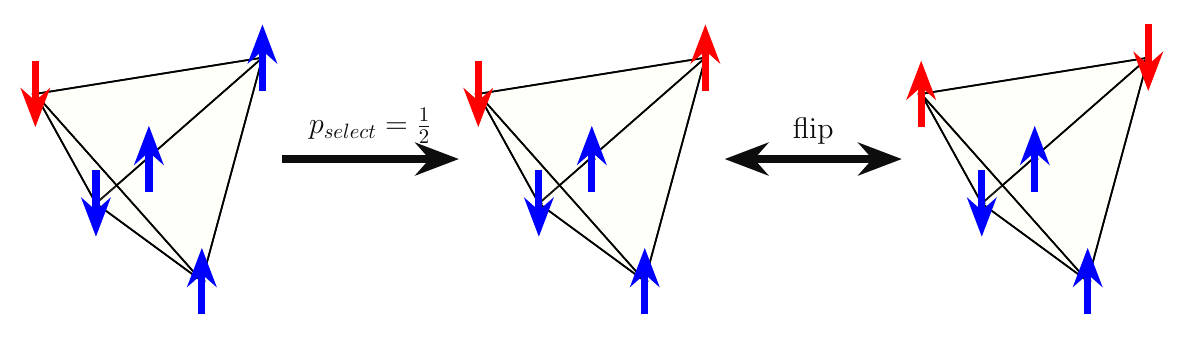}
    \end{subfigure}    
    \begin{subfigure}[b]{1.0\linewidth}
        \centering
        \includegraphics[width=\linewidth]{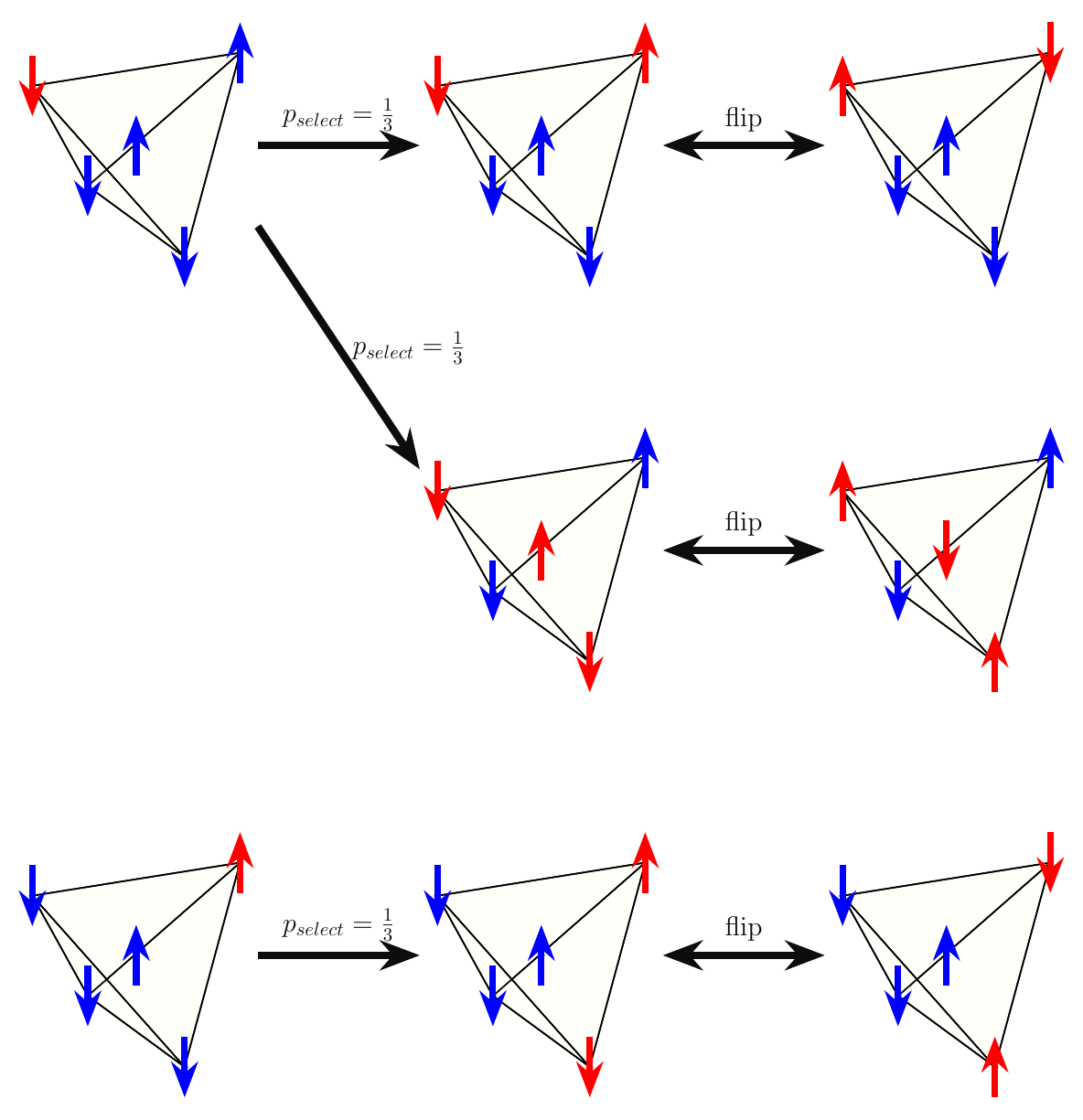}
    \end{subfigure}
    \caption{The update process on a single $2{:}2$- and $3{:}1{/}1{:}3, L$-tetrahedron satisfying the Gauss law. Spins participating in the loop are shown in red. In case of a new loop, select a random vertex spin, say the down spin in the upper left corner.  In case of an ongoing loop construction, imagine the loop entering the tetrahedron in the upper left corner.  Upper ($2{:}2$): Select with probability $1/2$ one of the two vertex spins of opposite orientation (up), and proceed to the adjacent tetrahedron of the selected spin. Repeat until the initial position is reached, in which case all spins along the loop are flipped. Middle ($3{:}1{/}1{:}3, L$ -- case of initial spin being of the majority type): Select one of the three vertex spins randomly with probability $1/3$. If the selected spin has the same orientation as the initial spin, add the center spin to the loop. Proceed to the adjacent tetrahedron of the selected spin and repeat until the initial position is reached, in which case all spins along the loop are flipped. Lower ($3{:}1{/}1{:}3, L$ -- case of initial spin being of the minority type ): Select one of the three vertex spins randomly with probability $1/3$ and flip together with the initial spin. }
    \label{fig:tetrahedra in loop update}
\end{figure}

As discussed previously, our analysis is valid at low temperatures, as long as 4:0/0:4 states can be ignored. In order to test this hypothesis, we perform classical Monte Carlo simulations for a system of linear length $L = 6$, with $6\times6\times6$ conventional unit cells, and $N = 24L^3 = 5184$ spins, subject to periodic boundary conditions. 

Two different algorithms are employed. \\
The first is the single spin-flip update with a Metropolis acceptance factor~\cite {Metropolis1953Equation}, which allows configurations to switch between different energy manifolds. Usually, one also applies annealing~\cite {kirkpatrick1983optimization}, which is the process of starting the simulation at a sufficiently high temperature and cooling it down step by step to a target temperature. This is heuristically known to enhance thermalization and prevent the configuration from being trapped in metastable states that are higher than the global minimum. Fortunately,  we did not encounter such metastabilities for the Ising model on the CPy lattice. However, the entropy of the ground state in the liquid phase is extensive~\cite{NutakkiScipost2023}, and the single spin-flip Metropolis algorithm has difficulties exploring all allowed configurations efficiently. An annealing procedure does not help, so we decided against using annealing.

The second type is a non-local, loop-update algorithm. We note that various kinds of loop updates~\cite{PhysRevE.57.1155,Roger_G_Melko_2004} have been developed for dipolar spin ice system before~\cite{Gingras2011,doi:10.1142/S2010324715400056,PhysRevX.4.011007}. Such a non-local update is similar to the classical worm algorithm~\cite{PhysRevLett.87.160601}, and is sometimes given the same name~\cite{jaubert2009topological,PhysRevX.4.011007}.  It will be sufficient for our purposes to only consider closed loops. 
The idea of a loop update is to temporarily break the Gauss law on a tetrahedron by introducing a defect ({\it i.e.}, flipping a spin for an Ising model) and then move around this defect over configuration space until all Gauss violations are healed, cf. Fig.\ref{fig:tetrahedra in loop update}. This leaves the energy of the final configuration equal to that of the initial configuration, and the updating process is its own reverse.
\begin{figure}
    \centering
    \includegraphics[width=1.0\linewidth]{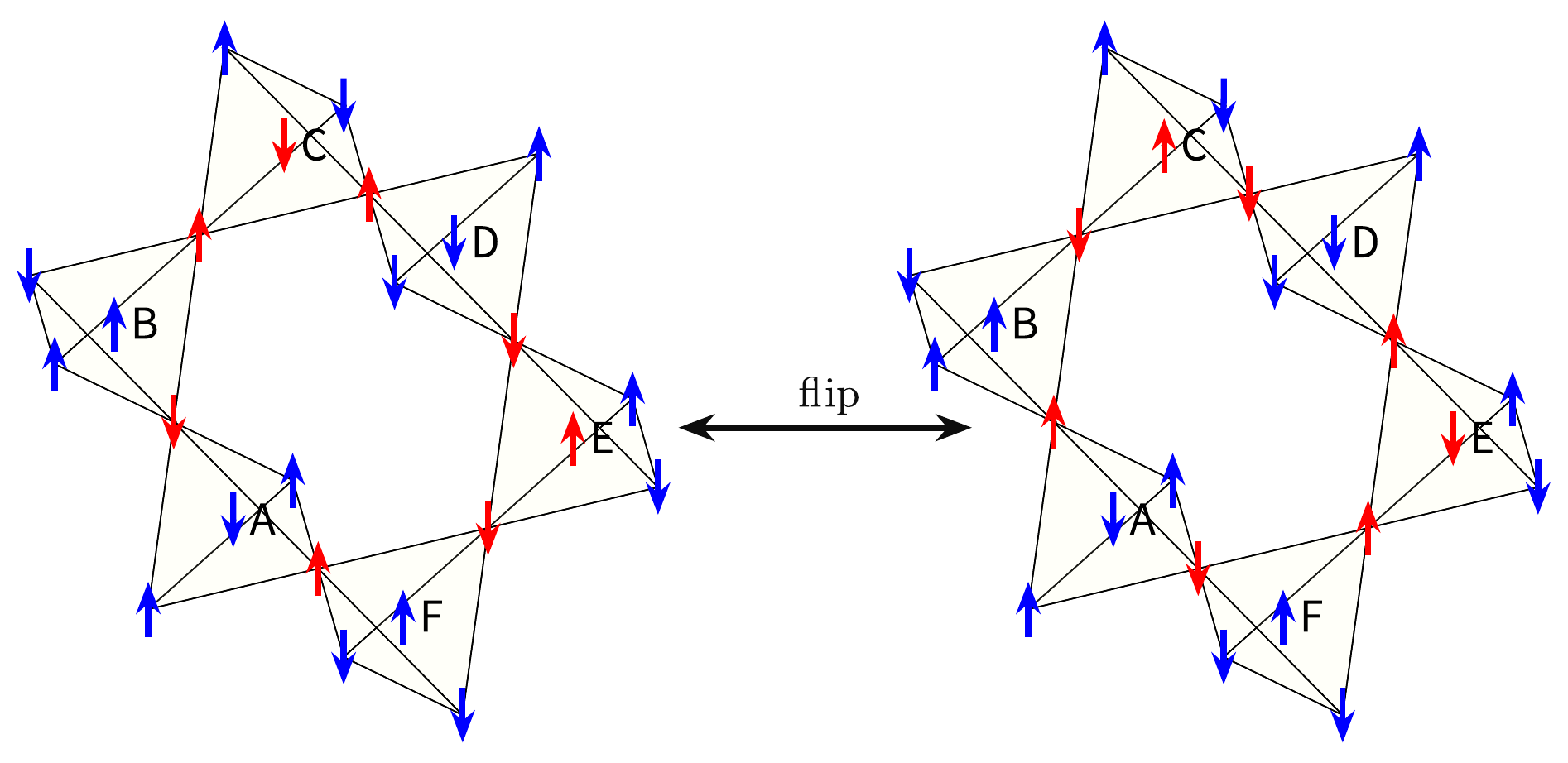}
    \caption{Construction of a full loop, consisting of red spins.  The $A$, $C$, $E$, $F$ tetrahedra belong to $3:1/1:3,L$ manifold, while $B$,$D$ belong to the $2:2$ manifold. The number of spins participating in the loop update is always even, and they are always arranged alternatingly upward and downward. The update is its own reverse.}
    \label{fig:shortest_loop_tetrahedra}
\end{figure}
In CPy, our focus is on both $2:2$-tetrahedra and $ 1:3/3:1; L$-tetrahedra. For constructing the loop update, both types of tetrahedra should be taken into account. After the loop update, the type of tetrahedra involved in the non-local update should be kept the same (see Fig.\ref{fig:shortest_loop_tetrahedra}), in order to keep the energy unchanged. The spin-selection process for $2:2$-tetrahedra should therefore be the one as in Fig.\ref{fig:tetrahedra in loop update}, upper panel; for $3:1/1:3, L$-tetrahedra, the selection process is not difficult either: We start by randomly selecting a vertex spin, and then choose one of the three other vertices at the same tetrahedron randomly with probability  $p_{\rm sel} = \frac{1}{3}$. If the selected spin has the same orientation as the initial spin, we take the center spin into the loop, which is guaranteed to have the opposite direction to the two vertex spins on a Gauss's law satisfying tetrahedron (see Fig.~\ref{fig:tetrahedra in loop update}, lower panel). After that, we proceed to the adjacent tetrahedron on the selected spin, where this procedure is repeated until the initial position is reached again.
Fig.~\ref{fig:shortest_loop_tetrahedra} shows an example of the loop update. In the $6\times6\times6$ system, the shortest closed loop length is $6$. 

We alternate between a full lattice sweep of $N = 24L^3 = 5184$ single spin-flip updates and loop updates, ensuring ergodicity of the sampling. We typically perform $10^5$ thermalization sweeps and $2\times10^5$ measurement sweeps. Zero temperature is approximated by setting $T_{{\rm MC}} = 0.001$, in which case the weight of the higher energy manifold is negligible. 


\section{\label{sec:Results}The spin structure factor}
The spin structure factor can be measured experimentally by using neutron diffraction. To facilitate the analysis, we split it into three parts: the vertex-vertex spin structure factor $S_{vv}(\mathbf{k}; x)$, the center-center spin structure factor $S_{cc}(\mathbf{k}; x)$, and the center-vertex spin structure factor $S_{cv}(\mathbf{k}; x)$ -- a similar split was already seen explicitly in the analytically derived expressions for the spin correlator.

\begin{figure*}[p]
    \centering
    \begin{subfigure}[b]{0.9\linewidth} 
        \centering
        \includegraphics[width=\linewidth]{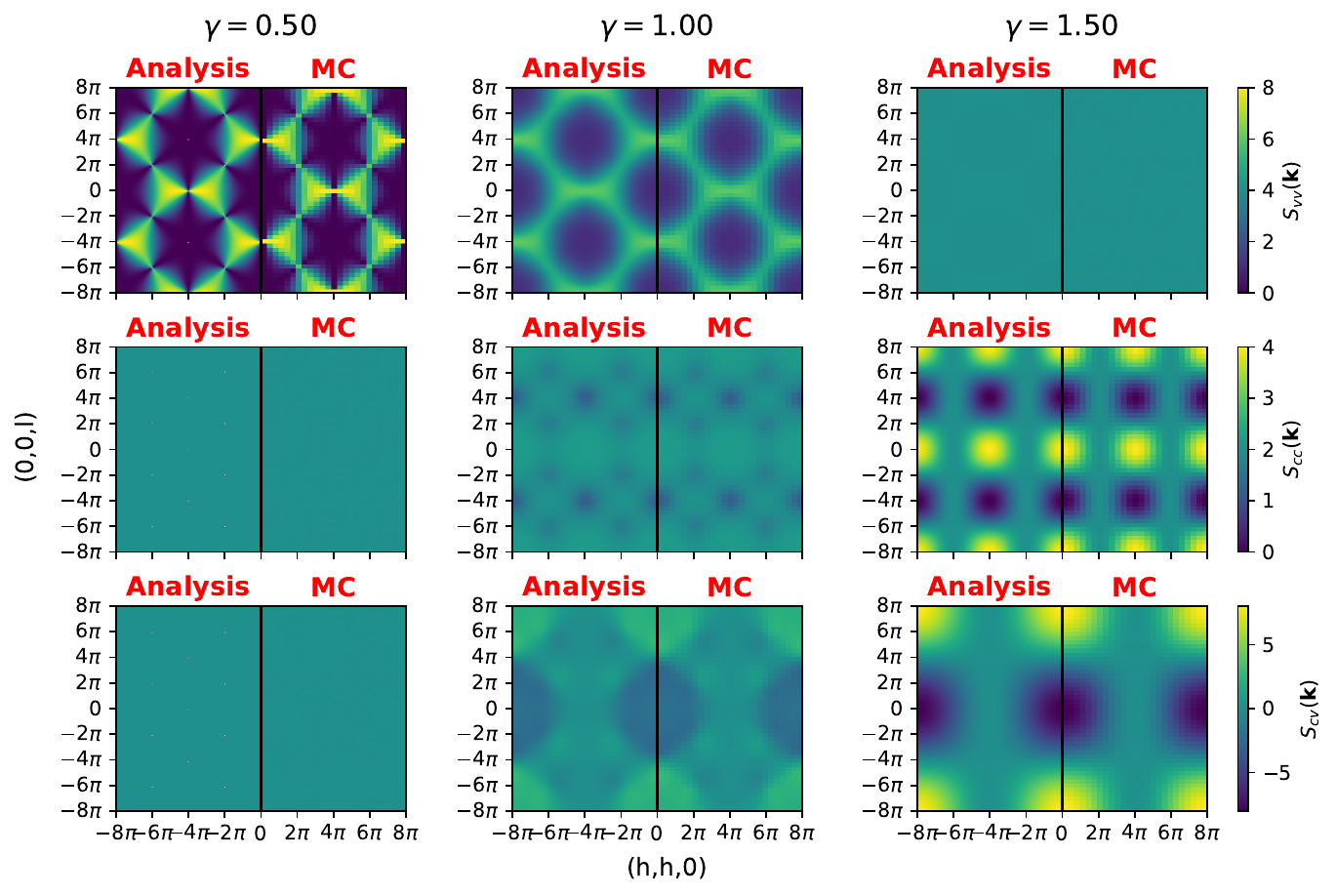}
        \caption{}
     \label{subfig:Figs/Both_Analysis_Numerics_FT_three_gammas_T=0.0100_2.pdf}
    \end{subfigure}    
    \begin{subfigure}[b]{0.9\linewidth}
        \centering
        \includegraphics[width=\linewidth]{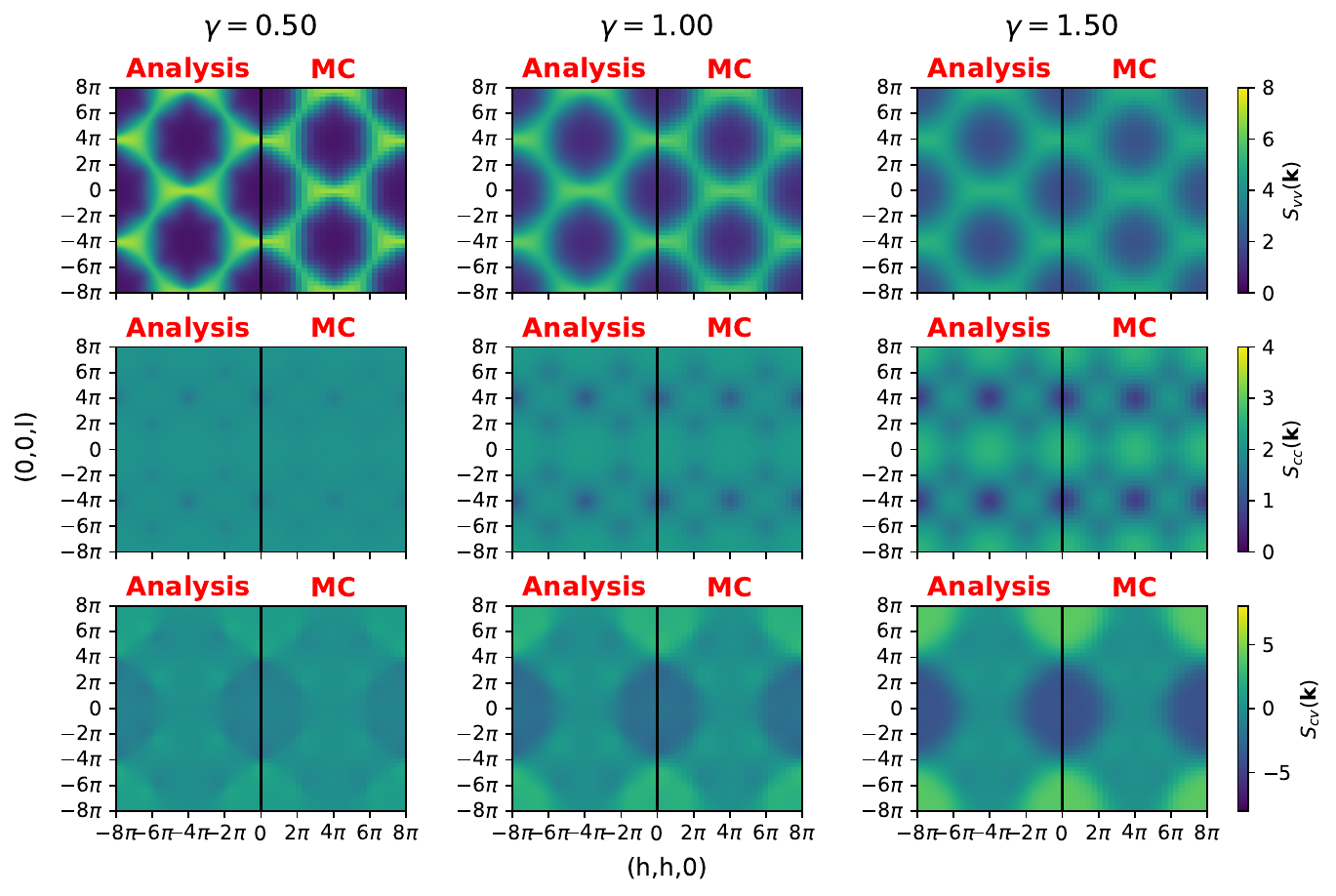}
        \caption{}
        \label{subfig:Figs/Both_Analysis_Numerics_FT_three_gammas_T=1.0000_2.pdf}
    \end{subfigure}
    \caption{The three contributions to the structure factor calculated from the Gaussian approximation (left) and the MC simulations (right) for three different phases ($\gamma = 0.50$ : spin ice, $\gamma = 1.00$ : coexistence, and $\gamma = 1.50$ : monopole fluid) at $T = 0.00$((a), upper panel) and $T = 1.00$((b), lower panel).}
    \label{fig:HHL}
\end{figure*}

\begin{figure*}[p]
  \centering
  
  \begin{subfigure}[b]{0.45\textwidth}
    \centering
    \includegraphics[width=\textwidth]{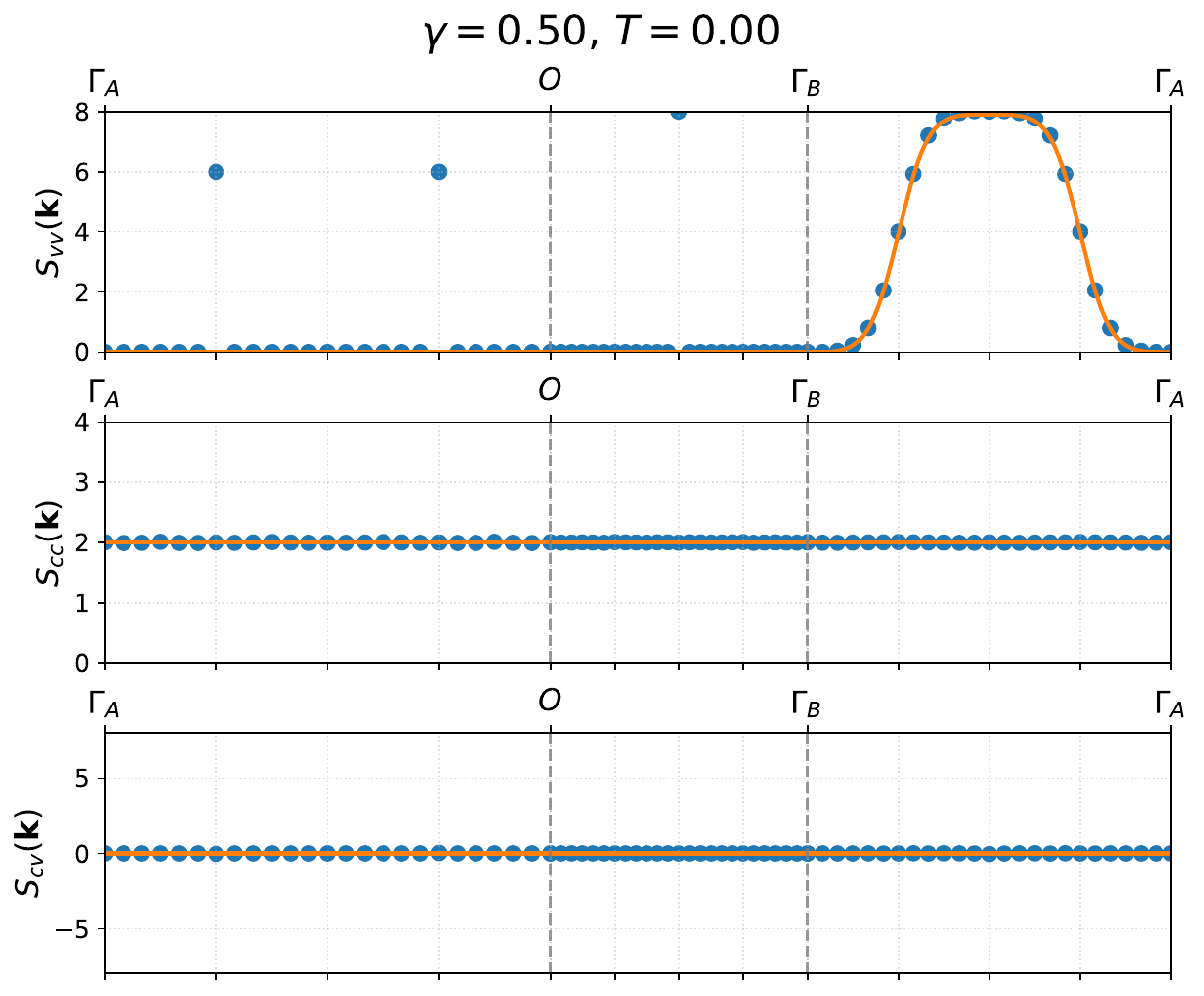}
    \caption{}
    \label{subfig:Figs/Tri_Fitting_Gaussian_Gamma=0.50_GaussianGamma=2.00_T=0.0010.pdf}
  \end{subfigure}
  \hfill 
  \begin{subfigure}[b]{0.45\textwidth}
    \centering
    \includegraphics[width=\textwidth]{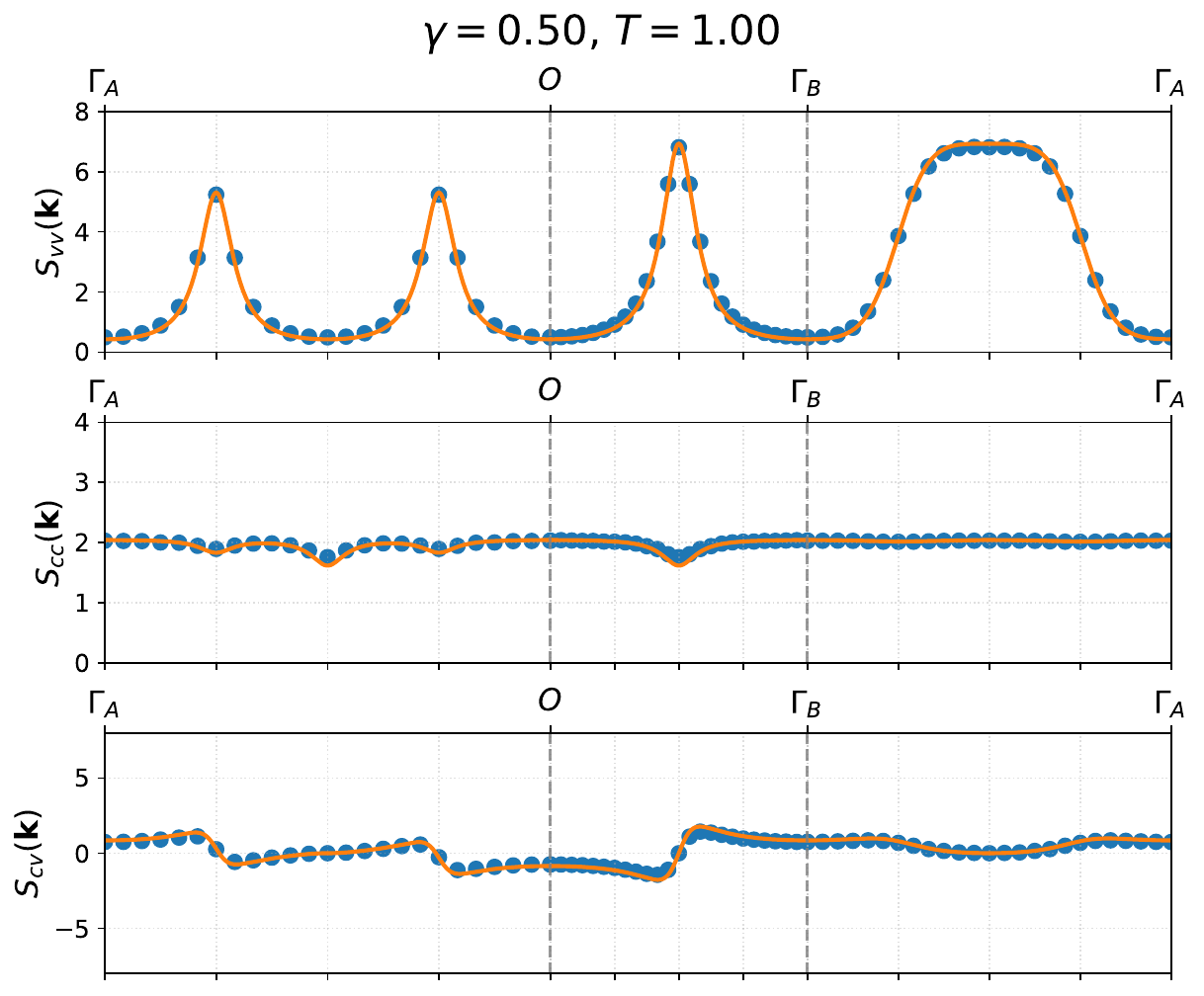}
    \caption{}
    \label{subfig:Figs/Tri_Fitting_Gaussian_Gamma=0.50_GaussianGamma=2.00_T=1.0000.pdf}
  \end{subfigure}

  \vspace{0cm} 

  \begin{subfigure}[b]{0.45\textwidth}
    \centering
    \includegraphics[width=\textwidth]{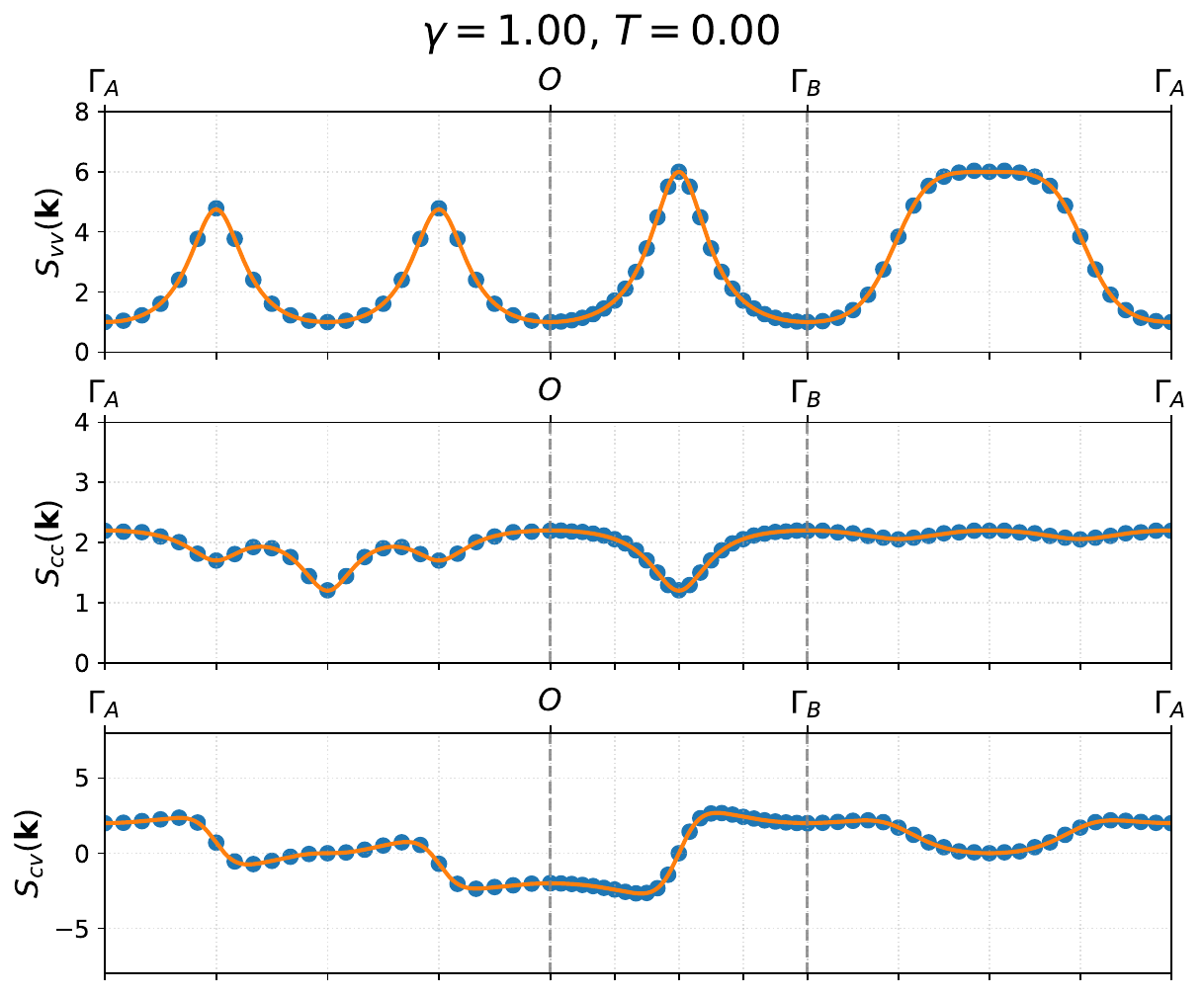}
    \caption{}
    \label{subfig:Figs/Tri_Fitting_Gaussian_Gamma=1.00_GaussianGamma=2.00_T=0.0010.pdf}
  \end{subfigure}
  \hfill
  \begin{subfigure}[b]{0.45\textwidth}
    \centering
    \includegraphics[width=\textwidth]{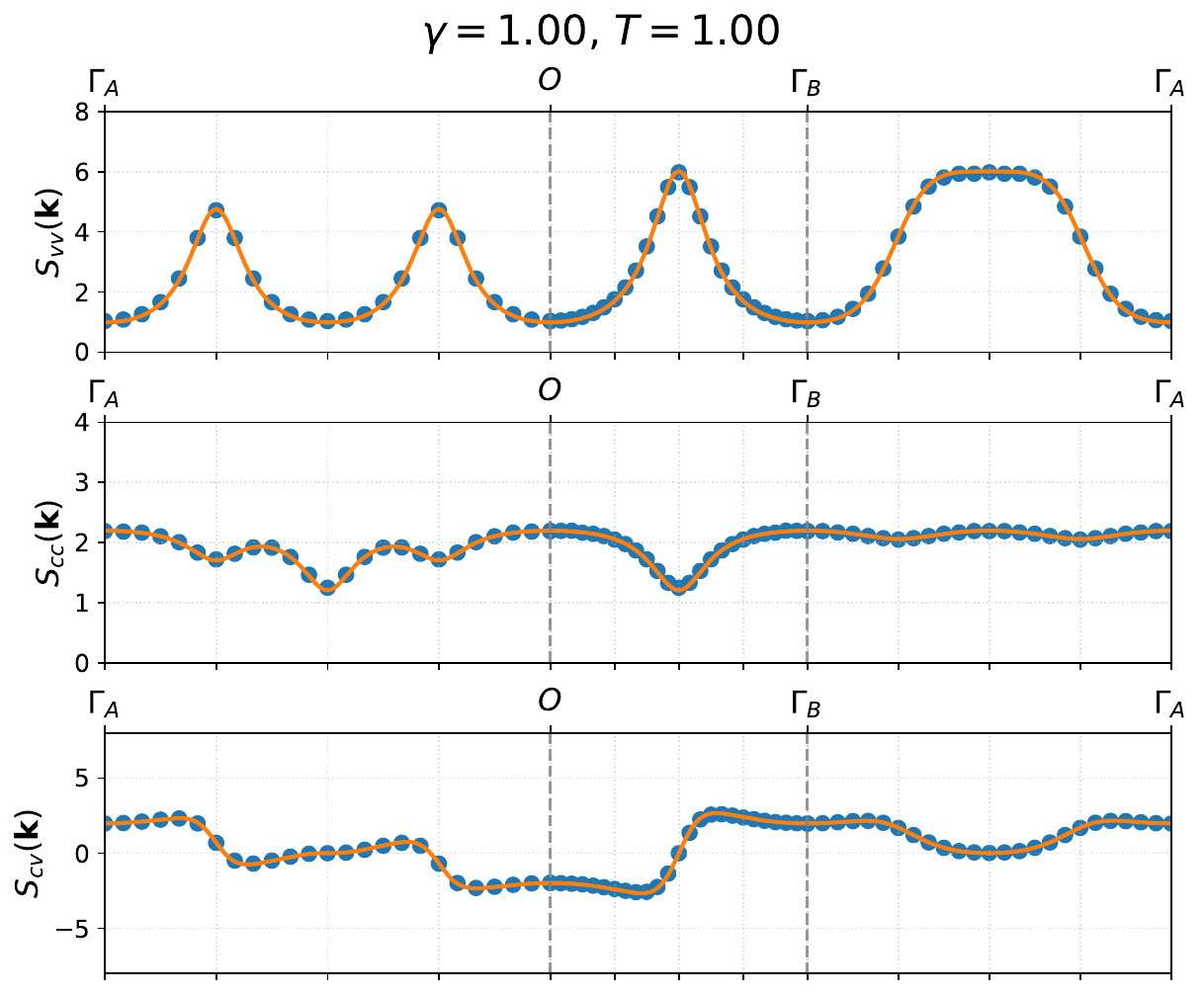}
    \caption{}
    \label{subfig:Figs/Tri_Fitting_Gaussian_Gamma=1.00_GaussianGamma=2.00_T=1.0000.pdf}
  \end{subfigure}

  \vspace{0cm}

  \begin{subfigure}[b]{0.45\textwidth}
    \centering
    \includegraphics[width=\textwidth]{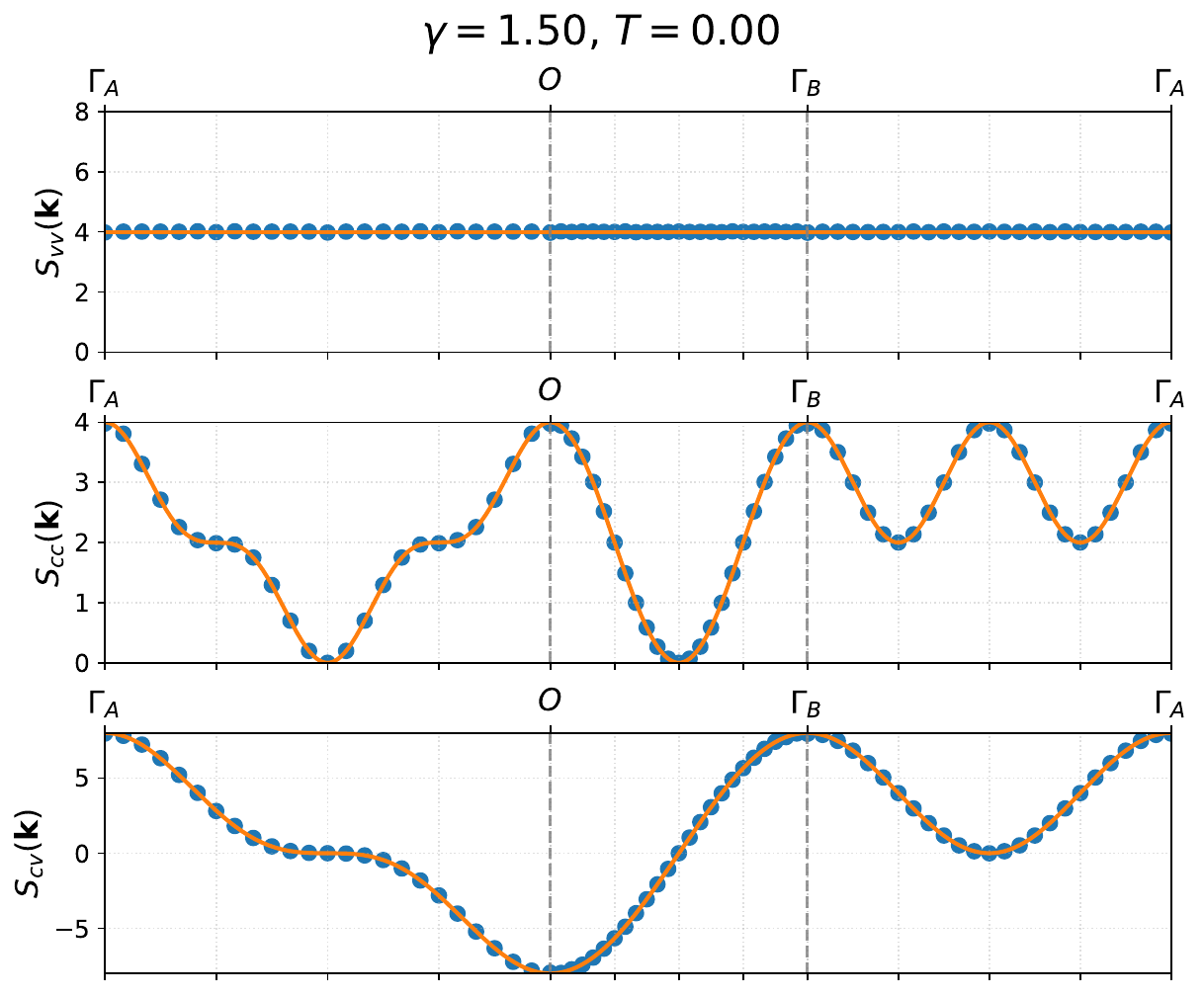}
    \caption{}
    \label{subfig:Figs/Tri_Fitting_Gaussian_Gamma=1.50_GaussianGamma=2.00_T=0.0010.pdf}
  \end{subfigure}
  \hfill
  \begin{subfigure}[b]{0.45\textwidth}
    \centering
    \includegraphics[width=\textwidth]{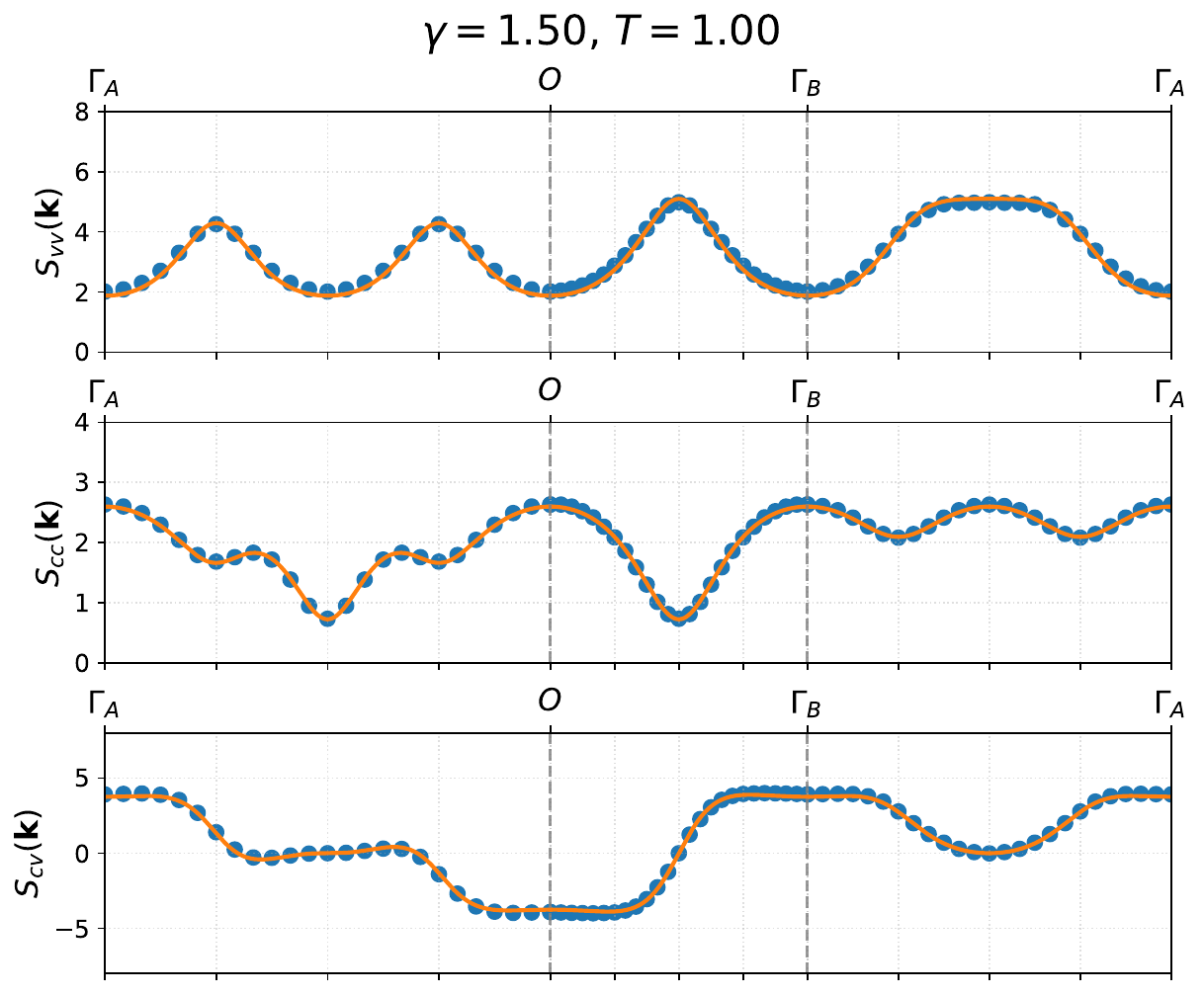}
    \caption{}
    \label{subfig:Figs/Tri_Fitting_Gaussian_Gamma=1.50_GaussianGamma=2.00_T=1.0000.pdf}
  \end{subfigure}

  \caption{Comparison of the analytical results (orange line) and the MC results (blue points) along a triangular path across the Brillouin zone: $\Gamma_{A} = 2\pi(-4, -4,4)\rightarrow O = 2\pi(0,0,0)\rightarrow\Gamma_{B} = 2\pi(0,0,4)\rightarrow\Gamma_{A} = 2\pi(-4,-4,4)$ in the $HHL$-plane.  The parameters are the same as in Fig.\ref{fig:HHL}. The analytical and numerical results match virtually exactly without any fitting. }
  \label{fig:Tri}
\end{figure*}

For each of the three contributions, we separately compare the predictions of the analytical theory with the statistically exact Monte Carlo results. Excellent agreement is found over the full validity range of the analytical theory: this is seen not only in the color plot of Fig.~\ref{fig:HHL} but also along a specific momentum trajectory in the Brillouin zone shown in  Fig.~\ref{fig:Tri}.
 
For spin ice, $0<\gamma <1$, we refer to the left column of Fig.~\ref{fig:HHL} and the upper row of Fig.~\ref{fig:Tri}. In Fig.~\ref{fig:HHL}, a clear pinch point in $S_{vv}(\mathbf{k})$ is seen, and it is the same one as in Ref.~\cite{PhysRevB.71.014424}. More pinch points are shown in Fig.~\ref{fig:Tri} with weight 8 and 6, respectively. The $S_{cv}(\mathbf{k}) \equiv 0$ and $S_{cc}(\mathbf{k})\equiv 2$ structure factors are flat,  demonstrating that center spins fully decorrelate from vertex spins and that center spins have only self-correlations.
 
For the monopole phase, $1 < \gamma < 3$, we refer to the right column of Fig.~\ref{fig:HHL} and the lower row of Fig.~\ref{fig:Tri}. Now $S_{vv}(\mathbf{k}) \equiv4$, showing that the vertex spins only have self-correlations while $S_{cv}$ and $S_{cc}$ exhibit oscillatory behavior, which arises from the fact that only nearest-neighbor correlations exist. 

For the point $\gamma = 1$, which contains both $2{:}2$-tetrahedra and $3{:}1{/}1{:}3; L$-tetrahedra, we refer to the middle column of Fig.~\ref{fig:HHL} and the middle row of Fig.~\ref{fig:Tri}.  The $S_{vv}(\mathbf{k})$ structure factor shows broadened pinch points which were originally at $2\pi(0,0,2)$ and $2\pi(-1,-1,1)$. This broadening is due to the fact that this phase is not strictly a Coulomb phase because the $3{:}1{/}1{:}3;L$-tetrahedra act as fluctuating topological charge sources, while the $2{:}2$-tetrahedra are electrically neutral.

Finally, we extrapolate our methods to finite temperature. At low finite temperature, when $\gamma$ is close to $1$ (in Fig.\ref{fig:HHL} and Fig.\ref{fig:Tri}, we take $0.5\leq\gamma \leq 1.5$), we can still assume the system is a two-level ensemble similar to the $\gamma = 1$ case. The $S_{vv}(\mathbf{k})$ structure factors characteristically show a broadened pinch point for all values of $\gamma$. $S_{cc}(\mathbf{k})$, $S_{cv}(\mathbf{k})$ have a similar behavior. The width of the broadened pinch point only depends on the concentration of $3{:}1{/}1{:}3; L$-tetrahedra, which itself is given by a Boltzmann occupation number, cf. Fig.~\ref{fig:Figs/CPy_n_gamma_T.pdf}. 

\begin{figure}
    \centering
    \includegraphics[width=1.0\linewidth]{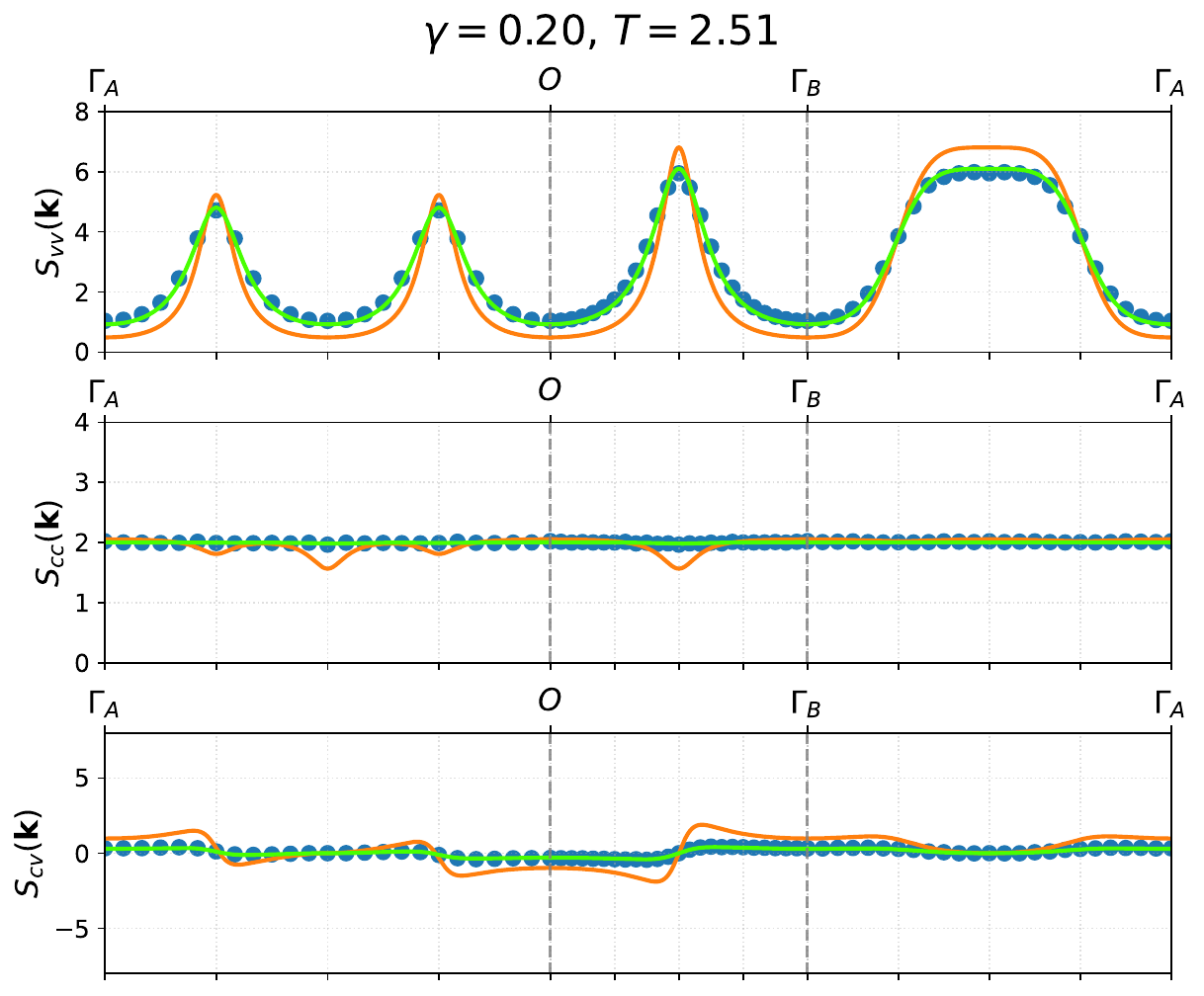}
    \caption{Comparison of the two-level ensemble assumption (orange line), the three-level ensemble assumption (green line) and the MC results (blue points) along a triangular path across the Brillouin zone: $\Gamma_{A} = 2\pi(-4, -4,4)\rightarrow O = 2\pi(0,0,0)\rightarrow\Gamma_{B} = 2\pi(0,0,4)\rightarrow\Gamma_{A} = 2\pi(-4,-4,4)$ in the $HHL$-plane.  The analytical and numerical results match virtually exactly without any fitting. }
    \label{fig:Tri_2level_VS_3level}
\end{figure}

For small $\gamma$ ($0 < \gamma \leq 0.5$), the energy gap between $3{:}1{/}1{:}3; L$-tetrahedra and $3{:}1{/}1{:}3; H$-tetrahedra is not negligible (see Fig.~\ref{fig:energy_levels}). The three-level ensemble provides then quantitatively more accurate results at elevated temperatures, as can be seen in Fig.~\ref{fig:Tri_2level_VS_3level}. Conceptually it enables to include in the Gaussian formalism, that was initially designed for ground-state physics, states that are never part of the ground states. As a consequence one can explore analytically and accurately what happens when monopoles in spin ice take two different flavors.



\section{Conclusion and Further Discussion}
\label{chapter:Discussion}

In conclusion, we computed the spin correlation functions of the liquid-like phases of the $J_1 -J_2$ Ising model on the CPy lattice analytically, and found the results to be in excellent agreement with Monte Carlo simulations. Exact results for the spin correlation functions were found in the monopole fluid $(1 < \gamma < 3)$ ground state. Because the center spins are free to fluctuate in the spin ice phase for independent tetrahedra, the ground-state correlations between the vertex and center spins vanish, and the ground-state vertex-vertex spin correlations are the same ones as on the pyrochlore lattice, which we computed using Henley's method.  At $\gamma =1$ and at finite temperature, we extended Henley's method to statistical mixtures with inequivalent center and vertex spins. Remarkably, for $\gamma \ll 1$ we could extend the description to a 3-level system, thereby extending the temperature range of the analytical approach to $T \sim 3J$.

The present results are easily transferable to other models where the monopole fluid competes with the 2:2 Coulomb phase on pyrochlore, e.g. in models with four-body interactions \cite{doi:10.1142/S2010324715400056,Slobinsky18a}, or possibly magneto-elastic coupling \cite{Slobinsky21a}. Our extension of Henley's method is as well implementable in other centered Ising models (e.g. on the kagome or checkerboard lattices), Heisenberg spins, and should also have applicability beyond equilibrium settings.


\begin{acknowledgments}
We acknowledge financial support from the Deutsche Forschungsgemeinschaft (DFG), project Nr 530111096.
The ALPSCore libraries were used in the Monte Carlo simulations~\cite{ALPSCore1,ALPSCore2}.
The derivations and calculations provided in this work were done prior to the availability of performant AI tools; nevertheless, we checked and sharpened the derivations by making use of ChatGPT 5.5. The AI found a great simplification for the three-level system. The text was exclusively written by humans.
\end{acknowledgments}

\appendix

\section{Lattice Definition}
\label{Appendix:Lattice Definition}

The centered pyrochlore(CPy) lattice is the combination of the pyrochlore lattice consisting of shared corner tetrahedra arranged in fcc symmetry, and its own mediate(dual) diamond lattice\cite{PhysRevB.71.014424} \cite{annurev:/content/journals/10.1146/annurev-conmatphys-070909-104138}, located at the center of each tetrahedron (see Fig.\ref{fig:CPy}). The position of the sites can be explicitly defined as
\begin{equation}
    \mathbf{r}_{I, \mu} = \mathbf{R}_{I} + \bm{\delta}_{{\mu}} \text{,}
    \label{eq:II.A.1}
\end{equation}
where $\mathbf{R}_I = \sum_{i=1}^{3} n_i \mathbf{a}_i$ representing the position of fcc primitive cells, with $\mathbf{a}_1 = \frac{1}{2}(0,1,1)^T$, $\mathbf{a}_2 = \frac{1}{2}(1,0,1)^T$, $\mathbf{a}_3 =\frac{1}{2}(1,1,0)^T$. The sublattices can be written as 
\begin{gather}
    \bm{\delta}_{a}  = 
    \begin{pmatrix}
        0 ,0 ,0
    \end{pmatrix}^T\text{,} \\
    \bm{\delta}_b = 
    \frac{1}{4}\begin{pmatrix}
        1, 1, 1
    \end{pmatrix}^T\text{,} \\
    \bm{\delta}_{0}  = 
    \frac{1}{8}\begin{pmatrix}
        1 , 1 , 1
    \end{pmatrix}^{T}\text{,}\\
    \bm{\delta}_{1}  = 
    \frac{1}{8}\begin{pmatrix}
        1 , -1 , -1
    \end{pmatrix}^{T}\text{,}\\ 
    \bm{\delta}_{2}  = 
    \frac{1}{8}\begin{pmatrix}
        -1 , 1 , -1
    \end{pmatrix}^{T}\text{,} \\
    \bm{\delta}_{3}  = 
    \frac{1}{8}\begin{pmatrix}
        -1 , -1 , 1
    \end{pmatrix}^{T}\text{.}
\end{gather} \label{eq:I.A.2}
\\
In these units, the lattice constant of the conventional cubic unit cell is set to be $a=1$. Vertex sites have subindices $\mu = 0, 1, 2, 3$, while $\mu = a, b$ label the center spins forming the bipartite diamond lattice. They also belong to the $ a$- or $ b$-type of tetrahedra.


\section{Statistical Mechanics of a single tetrahedron}
\label{App:Stat_Mech}
In this appendix, we provide the statistical-mechanics description of a single tetrahedron. The main purpose is to compute the equilibrium density of excitations.
To set the stage, we consider the $2{:}2$ tetrahedra as the ground-state manifold. The other cases can be dealt with in a similar fashion. 

From the table in Sec.~\ref{table:GS}, we take the energy of each defect, and compute the energy cost for the 4 excited levels,

\begin{align}
    \Delta_{1+} &= \frac{J}{2}(2 - \gamma)^2 - \frac{J}{2}\gamma^2 = J(2 -2 \gamma),\\
    \Delta_{1-} &= \frac{J}{2}(2 + \gamma)^2 - \frac{J}{2}\gamma^2 = J(2 +2 \gamma),\\
    \Delta_{2+} &= \frac{J}{2}(4 - \gamma)^2 - \frac{J}{2}\gamma^2 = 2J(4 -2 \gamma),\\
    \Delta_{2-} &= \frac{J}{2}(4 + \gamma)^2 - \frac{J}{2}\gamma^2 = 2J(4 +2 \gamma),
\end{align}
where the subindices $1$ and $2$ label the absolute number of topological charge units in the $3{:}1{/}1{:}3$ and $4{:}0{/}0{:}4$ manifolds. The second subindex $\pm$ labels the center spin (anti-)aligning with the direction of the minority of the vertex spins, corresponding to the low($L$) and high($H$) energy manifolds, $\Delta_{1,2+} < \Delta_{1,2-}$.

The internal energy of this system is given by
\begin{multline}
    U = N_{1+}\Delta_{1+} + N_{1-} \Delta_{1-} + N_{2+}\Delta_{2+} + N_{2-}\Delta_{2-}\\
    =N_{t}(n_{1+}\Delta_{1+} + n_{1-} \Delta_{1-} + n_{2+}\Delta_{2+} + n_{2-}\Delta_{2-}) ,
\end{multline}
where $N_t$ is the total number of tetrahedra,  $N_{1,2\pm}$ the number of tetrahedra of the type specified by their subindices, and $n_{1,2\pm} \equiv N_{1,2\pm}/N_t$ the corresponding density. 

We compute the entropy of the CPy Ising system using Ryzhkin's method~\cite{article}. 
In a single tetrahedron, the number of different spin configurations is $2^{4+1} = 32$. According to their degeneracies which are shown in table in Sec.~\ref{table:GS}, we enumerate them as  $N_1 = N_2 = N_3 = \dots =N_{12} = (1 - n_{1+}-n_{1-}-n_{2+}-n_{2-})N_{t}/12$, $N_{13} = \dots =N_{20} = n_{1+}N_t/8$, $N_{21}=\dots =N_{28} = n_{1-}N_t/8$, $N_{29} = N_{30} = n_{2+}N_t/2$, and $N_{31} = N_{32} = n_{2-}N_t/2$.
The number of states for a particular realization of $(N_1, \ldots N_{32})$, treating the $N_j$ as independent, is then 
\begin{equation}
    W = \left( \frac{1}{2} \right)^{2N_t} \frac{N_t!}{N_1!N_2!\dots N_{32}!}.
\end{equation}
Note that the number of vertex spins is $\frac{4}{2}N_t = 2N_t$ and the number of center spins is $N_t$. Since every vertex spin participates in two tetrahedra, the prefactor $(\frac{1}{2})^{2N_t}$ was added to $W$.

The configurational entropy is then 
\begin{multline}
    S = \ln W = N_t \left[ -2\ln 2  \right. \\- (1 - n_{1+}-n_{1-}-n_{2+}-n_{2-})\ln\frac{1 - n_{1+}-n_{1-}-n_{2+}-n_{2-}}{12} \\ \left. - n_{1+}\ln\frac{n_{1+}}{8}- n_{1-}\ln\frac{n_{1-}}{8} - n_{2+}\ln\frac{n_{2+}}{2}-n_{2-}\ln\frac{n_{2-}}{2} \right].
\end{multline}

Minimizing the free energy $F = U -TS$ leads to the Boltzmann expressions for  $n_{1\pm, 2\pm}$:
\begin{eqnarray}
n_{1\pm} &= & 8 \exp\left(-\frac{\Delta_{1\pm}}{T}\right)/Z_1 \label{eq:Boltzmann1},\\
n_{2\pm } &= & 2 \exp\left(-\frac{\Delta_{2\pm}}{T}\right)/Z_1 \label{eq:Boltzmann2},\\
n_0 & = & 1 -  n_{1+}-n_{1-}-n_{2+}-n_{2-} = \frac{12}{Z_1} \label{eq:Boltzmann0},\\
Z_1 &= & 12 + 8\exp\left(-\frac{\Delta_{1+}}{T}\right)+ 8\exp\left(-\frac{\Delta_{1-}}{T}\right) \nonumber \\
{} & {}  &   + 2\exp\left(-\frac{\Delta_{2+}}{T}\right)+2\exp\left(-\frac{\Delta_{2-}}{T}\right)\textbf{.}
\end{eqnarray}

\subsubsection{Three level system}
We consider the statistical mechanics of the three level system with the $2{:}2$ spin ice ground state.
The excitation energies of the charged tetrahedra $3{:}1/1{:}3,L$ and $3{:}1/1{:}3,H$ are, respectively,
\begin{equation}
\Delta_L=2J(1-\gamma)\text{,}\qquad \Delta_H=2J(1+\gamma)\text{.}
\end{equation}
The partition function and population density are given by
\begin{align}
Z_{1}^{(3)}&=12+8e^{-\beta\Delta_L}+8e^{-\beta\Delta_H}\text{,}\\
n_L&=\frac{8e^{-\beta\Delta_L}}{Z_{1}^{(3)}}\text{,}
\quad
n_H=\frac{8e^{-\beta\Delta_H}}{Z_{1}^{(3)}}\text{,}\\
n&=n_L+n_H\text{,}
\quad n_0=1-n\text{.}
\end{align}

\begin{widetext}
\section{Projector operator}
\label{Appendix:Projector operator}

In Sec.\ref{sec:Gauss}, we introduced the projector operator for the CPy lattice $\Pi(\mathbf{k}; \alpha)$, with $\alpha = 2x$. Here, we provide the detailed derivation of $\Pi(\mathbf{k}; \alpha)$, which has a matrix structure over the unit-cell indices, cf. Ref.~\cite{PhysRevB.71.014424}. Our notation uses a block matrix structure for the center spins ($2 \times 2$) and the vertex spins ($4 \times 4$).

According to Eq.\ref{eq:constraint_k}, we have
\begin{equation}
    E_c^\dagger E_c(\mathbf{k};\alpha) = \alpha^2I_{2\times2} + E^\dagger E(\mathbf{k}) = 
    \begin{pmatrix}
        \alpha^2 +4 & 4H^*(\mathbf{k})\\
        4H(\mathbf{k}) & \alpha^2 + 4
    \end{pmatrix},
\end{equation}
where
\begin{equation}
    H(\mathbf{k}) = \frac{1}{4}\sum_{i = 0}^3 e^{2i\mathbf{k}\cdot\bm{\delta}_{i}} = \cos\frac{k_x}{4}\cos\frac{k_y}{4}\cos\frac{k_z}{4} -i\sin\frac{k_x}{4}\sin\frac{k_y}{4}\sin\frac{k_z}{4},
    \label{eq:App.1.H}
\end{equation}
and $H^*(\mathbf{k}) = H(-\mathbf{k})$.
Then
\begin{multline}
    [E_c^\dagger E_c(\mathbf{k};\alpha)]^{-1} 
    = \frac{1}{4}\left[\left(1+\frac{\alpha^2}{4}\right)^2 - |H(\mathbf{k})|^2\right]^{-1}
    \begin{pmatrix}
        \frac{1}{4}\alpha^2 +1 & -H^*(\mathbf{k})\\
        H(\mathbf{k}) & \frac{1}{4}\alpha^2 + 1
    \end{pmatrix} \\
    =   \frac{1}{4}\frac{1 - |H(\mathbf{k})|^2}{(1+\frac{\alpha^2}{4})^2 - |H(\mathbf{k})|^2} \left[\frac{\alpha^2}{4}\frac{1}{1-|H(\mathbf{k})|^2}I_{2\times2}] + \frac{1}{1 - |H(\mathbf{k})|^2}
    \begin{pmatrix}
        1 & -H^*(\mathbf{k})\\
        -H(\mathbf{k}) & 1
    \end{pmatrix}
    \right] \\
    = \frac{1}{16}\frac{\alpha^2}{(1+\frac{\alpha^2}{4})^2 - |H(\mathbf{k})|^2}I_{2\times2} + \frac{1 - |H(\mathbf{k})|^2}{(1+\frac{\alpha^2}{4})^2 - |H(\mathbf{k})|^2}(E^\dagger E(\mathbf{k}))^{-1},
\end{multline}
with~\cite{PhysRevB.71.014424}
\begin{equation}
    (E^\dagger E(\mathbf{k}))^{-1} = \frac{1}{4}\frac{1}{1 - |H(\mathbf{k})|^2}
    \begin{pmatrix}
        1 & -H^*(\mathbf{k}) \\ 
        -H(\mathbf{k}) & 1
    \end{pmatrix}.
\end{equation}
Plugging this into the equation for the projector operator, we find
\begin{multline}
    \Pi(\mathbf{k};\alpha) = \frac{1}{(1+\frac{\alpha^2}{4})^2 - |H(\mathbf{k})|^2}
    \\
    \begin{pmatrix}
        D I_{2\times2}-\alpha^2(1 - |H(\mathbf{k})|^2)(E^\dagger E(\mathbf{k}))^{-1} & \frac{\alpha^3}{16}E^\dagger+\alpha(1 - |H(\mathbf{k})|^2)(E^\dagger E(\mathbf{k}))^{-1}E^{\dagger} \\
        \frac{\alpha^3}{16}E+\alpha(1 - |H(\mathbf{k})|^2)E(E^\dagger E(\mathbf{k}))^{-1} & D I_{4\times4} - \frac{\alpha^2}{16}EE^{\dagger}-(1 - |H(\mathbf{k})|^2)E(E^\dagger E(\mathbf{k}))^{-1}E^{\dagger}
    \end{pmatrix},
    \label{eq:App.1.projector operator}
\end{multline}
with the short-hand notation $D=(1+\frac{\alpha^2}{2} - |H(\mathbf{k})|^2)$. Since $|H(\mathbf{k})|^2 = \frac{1}{4}(1 + \cos\frac{k_x}{2}\cos\frac{k_y}{2} + \cos\frac{k_y}{2}\cos\frac{k_z}{2}+ \cos\frac{k_z}{2}\cos\frac{k_x}{2}) \leq 1$, we note that there can only be a singularity in $\Pi(\mathbf{k};\alpha)$ when $\alpha=0$.

We proceed by computing the trace of the projector operator, $ {\rm Tr}\,(\Pi(\mathbf{k}; \alpha))$. Summing over the diagonal elements of the center site submatrix gives
\begin{multline}
    \Pi_{aa}(\mathbf{k};\alpha)+\Pi_{bb}(\mathbf{k};\alpha) = \frac{1}{(1+\frac{\alpha^2}{4})^2 - |H(\mathbf{k})|^2}\left[2\left(1+\frac{\alpha^2}{2} - |H(\mathbf{k})|^2\right) - \alpha^2(1 - |H(\mathbf{k})|^2){\rm Tr}\,[(E^\dagger E(\mathbf{k}))^{-1}]\right] \\
    = \frac{1}{(1+\frac{\alpha^2}{4})^2 - |H(\mathbf{k})|^2}\left[2\left(1+\frac{\alpha^2}{2} - |H(\mathbf{k})|^2\right) - \alpha^2(1 - |H(\mathbf{k})|^2)\frac{1}{2}\frac{1}{1 - |H(\mathbf{k})|^2}\right] 
    = 2 - \frac{\frac{\alpha^2}{2}+\frac{\alpha^4}{8}}{(1+\frac{\alpha^2}{4})^2 - |H(\mathbf{k})|^2} \, ,
\end{multline}
while the sum over the diagonal elements of the vertex site submatrix gives
\begin{multline}
  \sum_{j=0}^3  \Pi_{jj}(\mathbf{k};\alpha)
    = \frac{1}{(1+\frac{\alpha^2}{4})^2 - |H(\mathbf{k})|^2}\left[4\times\frac{\alpha^2}{2} - \frac{\alpha^4}{16}{\rm Tr}\,[E^\dagger E(\mathbf{k})] +(1 - |H(\mathbf{k})|^2){\rm Tr}\,[\Pi_{\rm pyro}(\mathbf{k})]\right]  \\ = \frac{1}{(1+\frac{\alpha^2}{4})^2 - |H(\mathbf{k})|^2}\left[4\times\frac{\alpha^2}{2} - \frac{\alpha^4}{16}\times8 +(1 - |H(\mathbf{k})|^2)\times2\right] = 2 + \frac{\frac{\alpha^2}{2}+\frac{\alpha^4}{8}}{(1+\frac{\alpha^2}{4})^2 - |H(\mathbf{k})|^2} .
\end{multline}
Here, $\Pi_{\rm pyro}(\mathbf{k}) = I_{4\times4} - E(E^\dagger E(\mathbf{k}))^{-1}E^{\dagger}$ is the projector used for the pure pyrochlore lattice, with ${\rm Tr}\,[\Pi_{\rm pyro}(\mathbf{k})] \equiv 2, \forall\mathbf{k}$, in line with the conclusion from Ref.~\cite{PhysRevB.71.014424}. We now combine both sums. Clearly,
\begin{equation}
    {\rm Tr}\,[\Pi(\mathbf{k}; \alpha)] \equiv 4, \forall\mathbf{k},\, \alpha \, ,
\end{equation}
from which follows that, for non-zero $\alpha$,
\begin{equation}
    {\rm Tr}\,P(\mathbf{\alpha}) = \int_{\mathbf{k}\in1 {\rm BZ}}\frac{d^3\mathbf{k}}{V_{1BZ}}{\rm Tr}\, \Pi(\mathbf{k}; \alpha) \equiv 4.
\end{equation}

\end{widetext}


\bibliography{CPYrefs}

@article{Slobinsky18a,
  title = {Charge and spin correlations in the monopole liquid},
  author = {Slobinsky, D. and Baglietto, G. and Borzi, R. A.},
  journal = {Phys. Rev. B},
  volume = {97},
  issue = {17},
  pages = {174422},
  numpages = {11},
  year = {2018},
  month = {May},
  publisher = {American Physical Society},
  doi = {10.1103/PhysRevB.97.174422},
  url = {https://link.aps.org/doi/10.1103/PhysRevB.97.174422}
}

@article{slobinsky21a,
	title = {Monopole matter from magnetoelastic coupling in the {Ising} pyrochlore},
	volume = {4},
	copyright = {2021 The Author(s)},
	issn = {2399-3650},
	url = {http://www.nature.com/articles/s42005-021-00552-0},
	doi = {10.1038/s42005-021-00552-0},
	number = {1},
	journal = {Communications Physics},
	author = {Slobinsky, D. and Pili, L. and Baglietto, G. and Grigera, S. A. and Borzi, R. A.},
	month = mar,
	year = {2021},
	pages = {1--11}
}

@article{PhysRevB.71.014424,
  title = {Power-law spin correlations in pyrochlore antiferromagnets},
  author = {Henley, C. L.},
  journal = {Phys. Rev. B},
  volume = {71},
  issue = {1},
  pages = {014424},
  numpages = {11},
  year = {2005},
  month = {Jan},
  publisher = {American Physical Society},
  doi = {10.1103/PhysRevB.71.014424},
  url = {https://link.aps.org/doi/10.1103/PhysRevB.71.014424}
}

@Article{NutakkiScipost2023,
	title={{The classical Heisenberg model on the centred pyrochlore lattice}},
	author={Rajah P. Nutakki and Ludovic D. C. Jaubert and Lode Pollet},
	journal={SciPost Phys.},
	volume={15},
	pages={040},
	year={2023},
	publisher={SciPost},
	doi={10.21468/SciPostPhys.15.2.040},
	url={https://scipost.org/10.21468/SciPostPhys.15.2.040},
}

@article{annurev:/content/journals/10.1146/annurev-conmatphys-070909-104138,
   author = "Henley, Christopher L.",
   title = "The “Coulomb Phase” in Frustrated Systems", 
   journal= "Annual Review of Condensed Matter Physics",
   year = "2010",
   volume = "1",
   number = "Volume 1, 2010",
   pages = "179-210",
   doi = "https://doi.org/10.1146/annurev-conmatphys-070909-104138",
   url = "https://www.annualreviews.org/content/journals/10.1146/annurev-conmatphys-070909-104138",
   publisher = "Annual Reviews",
   issn = "1947-5462",
   type = "Journal Article",
  }

@article{Nutakki2025,
  title = {Coupling quantum spin ice to matter on the centered pyrochlore lattice},
  author = {Nutakki, Rajah P. and Capponi, Sylvain and Jaubert, Ludovic D. C. and Pollet, Lode},
  journal = {Phys. Rev. B},
  volume = {112},
  issue = {14},
  pages = {144437},
  numpages = {20},
  year = {2025},
  month = {Oct},
  publisher = {American Physical Society},
  doi = {10.1103/xrmm-vr6m},
  url = {https://link.aps.org/doi/10.1103/xrmm-vr6m}
}

@article{NutakkiPRR2023,
  title = {Frustration on a centered pyrochlore lattice in metal-organic frameworks},
  author = {Nutakki, Rajah P. and R\"o\ss{}-Ohlenroth, Richard and Volkmer, Dirk and Jesche, Anton and von Nidda, Hans-Albrecht Krug and Tsirlin, Alexander A. and Gegenwart, Philipp and Pollet, Lode and Jaubert, Ludovic D. C.},
  journal = {Phys. Rev. Res.},
  volume = {5},
  issue = {2},
  pages = {L022018},
  numpages = {6},
  year = {2023},
  month = {May},
  publisher = {American Physical Society},
  doi = {10.1103/PhysRevResearch.5.L022018},
  url = {https://link.aps.org/doi/10.1103/PhysRevResearch.5.L022018}
}

@article{https://doi.org/10.1002/cphc.200900873,
author = {Castelnovo, Claudio},
title = {Coulomb Physics in Spin Ice: From Magnetic Monopoles to Magnetic Currents},
journal = {ChemPhysChem},
volume = {11},
number = {3},
pages = {557-559},
doi = {https://doi.org/10.1002/cphc.200900873},
year = {2010}
}

@article{article,
author = {Ryzhkin, I.},
year = {2005},
month = {09},
pages = {481-486},
title = {Magnetic relaxation in rare-earth oxide pyrochlores},
volume = {101},
journal = {Journal of Experimental and Theoretical Physics},
doi = {10.1134/1.2103216}
}

@article{kirkpatrick1983optimization,
	author = {S. Kirkpatrick and C. D. Gelatt and M. P. Vecchi},
	doi = {10.1126/science.220.4598.671},
	journal = {Science},
	number = {4598},
	pages = {671–680},
	title = {Optimization by Simulated Annealing},
	volume = {220},
	year = {1983}
}

@Inbook{Gingras2011,
author="Gingras, Michel J. P.",
editor="Lacroix, Claudine
and Mendels, Philippe
and Mila, Fr{\'e}d{\'e}ric",
title="Spin Ice",
bookTitle="Introduction to Frustrated Magnetism: Materials, Experiments, Theory",
year="2011",
publisher="Springer Berlin Heidelberg",
address="Berlin, Heidelberg",
pages="293--329",
isbn="978-3-642-10589-0",
doi="10.1007/978-3-642-10589-0_12",
url="https://doi.org/10.1007/978-3-642-10589-0_12"
}

@article{Roger_G_Melko_2004,
doi = {10.1088/0953-8984/16/43/R02},
url = {https://doi.org/10.1088/0953-8984/16/43/R02},
year = {2004},
month = {oct},
publisher = {},
volume = {16},
number = {43},
pages = {R1277},
author = {Roger G Melko and Michel J P Gingras},
title = {Monte Carlo studies of the dipolar spin ice model},
journal = {Journal of Physics: Condensed Matter}
}

@article{PhysRevE.57.1155,
  title = {Monte Carlo simulation of ice models},
  author = {Barkema, G. T. and Newman, M. E. J.},
  journal = {Phys. Rev. E},
  volume = {57},
  issue = {1},
  pages = {1155--1166},
  numpages = {0},
  year = {1998},
  month = {Jan},
  publisher = {American Physical Society},
  doi = {10.1103/PhysRevE.57.1155},
  url = {https://link.aps.org/doi/10.1103/PhysRevE.57.1155}
}

@article{PhysRevLett.87.160601,
  title = {Worm Algorithms for Classical Statistical Models},
  author = {Prokof'ev, Nikolay and Svistunov, Boris},
  journal = {Phys. Rev. Lett.},
  volume = {87},
  issue = {16},
  pages = {160601},
  numpages = {4},
  year = {2001},
  month = {Sep},
  publisher = {American Physical Society},
  doi = {10.1103/PhysRevLett.87.160601},
  url = {https://link.aps.org/doi/10.1103/PhysRevLett.87.160601}
}

@article{doi:10.1142/S2010324715400056,
    author = {Jaubert, L. D. C.},
    title = {Monopole Holes in a Partially Ordered Spin Liquid},
    journal = {SPIN},
    volume = {05},
    number = {02},
    pages = {1540005},
    year = {2015},
    doi = {10.1142/S2010324715400056}
}

@article{Metropolis1953Equation,
    author = {Metropolis, Nicholas and Rosenbluth, Arianna W. and Rosenbluth, Marshall N. and Teller, Augusta H. and Teller, Edward},
    title = {Equation of State Calculations by Fast Computing Machines},
    journal = {The Journal of Chemical Physics},
    volume = {21},
    number = {6},
    pages = {1087-1092},
    year = {1953},
    month = {06},
    issn = {0021-9606},
    doi = {10.1063/1.1699114},
    url = {https://doi.org/10.1063/1.1699114},
}

@article{PhysRevX.4.011007,
  title = {Magnetic-Moment Fragmentation and Monopole Crystallization},
  author = {Brooks-Bartlett, M. E. and Banks, S. T. and Jaubert, L. D. C. and Harman-Clarke, A. and Holdsworth, P. C. W.},
  journal = {Phys. Rev. X},
  volume = {4},
  issue = {1},
  pages = {011007},
  numpages = {16},
  year = {2014},
  month = {Jan},
  publisher = {American Physical Society},
  doi = {10.1103/PhysRevX.4.011007},
  url = {https://link.aps.org/doi/10.1103/PhysRevX.4.011007}
}

@phdthesis{jaubert2009topological,
  title={Topological constraints and defects in spin ice},
  author={Jaubert, Ludovic DC},
  year={2009},
  school={Ecole normale sup{\'e}rieure de lyon-ENS LYON},
  url = {https://theses.hal.science/tel-00462970v1/file/Thesis.pdf}
}

@article{ALPSCore1,
        title = {Updated core libraries of the {ALPS} project},
        volume = {213},
        issn = {0010-4655},
        url = {https://www.sciencedirect.com/science/article/pii/S0010465516303885},
        doi = {10.1016/j.cpc.2016.12.009},
        urldate = {2022-09-21},
        journal = {Computer Physics Communications},
        author = {Gaenko, A. and Antipov, A. E. and Carcassi, G. and Chen, T. and Chen, X. and Dong, Q. and Gamper, L. and Gukelberger, J. and Igarashi, R. and Iskakov, S. and Könz, M. and LeBlanc, J. P. F. and Levy, R. and Ma, P. N. and Paki, J. E. and Shinaoka, H. and Todo, S. and Troyer, M. and Gull, E.},
        month = apr,
        year = {2017},
        pages = {235--251},
}

@techreport{ALPSCore2,
        title = {Updated {Core} {Libraries} of the {ALPS} {Project}},
        url = {http://arxiv.org/abs/1811.08331},
        number = {arXiv:1811.08331},
        urldate = {2022-09-21},
        institution = {arXiv},
        author = {Wallerberger, Markus and Iskakov, Sergei and Gaenko, Alexander and Kleinhenz, Joseph and Krivenko, Igor and Levy, Ryan and Li, Jia and Shinaoka, Hiroshi and Todo, Synge and Chen, Tianran and Chen, Xi and LeBlanc, James P. F. and Paki, Joseph E. and Terletska, Hanna and Troyer, Matthias and Gull, Emanuel},
        month = nov,
        year = {2018},
        doi = {10.48550/arXiv.1811.08331},
}

@article{harris1997,
  title = {Geometrical Frustration in the Ferromagnetic Pyrochlore {Ho$_2$Ti$_2$O$_7$}},
  author = {Harris, M. J. and Bramwell, S. T. and McMorrow, D. F. and Zeiske, T. and Godfrey, K. W.},
  year = {1997},
  month = sep,
  journal = {Phys. Rev. Lett.},
  volume = {79},
  number = {13},
  pages = {2554--2557},
  issn = {0031-9007, 1079-7114},
  doi = {10.1103/PhysRevLett.79.2554},
  urldate = {2022-04-12},
  langid = {english},
}

@book{udagawa2021,
  title = {Spin {{Ice}}},
  editor = {Udagawa, Masafumi and Jaubert, Ludovic},
  year = {2021},
  series = {Springer {{Series}} in {{Solid-State Sciences}}},
  volume = {197},
  publisher = {{Springer International Publishing}},
  address = {{Cham}},
  doi = {10.1007/978-3-030-70860-3},
  urldate = {2023-03-17},
  isbn = {978-3-030-70858-0 978-3-030-70860-3},
  langid = {english},
}

@article{castelnovo2007,
  title = {Topological Order and Topological Entropy in Classical Systems},
  author = {Castelnovo, Claudio and Chamon, Claudio},
  year = {2007},
  month = nov,
  journal = {Phys. Rev. B},
  volume = {76},
  number = {17},
  pages = {174416},
  issn = {1098-0121, 1550-235X},
  doi = {10.1103/PhysRevB.76.174416},
  urldate = {2023-06-01},
  langid = {english},
}

@article{gandara2012,
  title = {Porous, Conductive Metal-Triazolates and Their Structural Elucidation by the Charge-Flipping Method},
  author = {G{\'a}ndara, Felipe and {Uribe-Romo}, Fernando J. and Britt, David K. and Furukawa, Hiroyasu and Lei, Liao and Cheng, Rui and Duan, Xiangfeng and O'Keeffe, Michael and Yaghi, Omar M.},
  year = {2012},
  month = aug,
  journal = {Chem. Eur. J.},
  volume = {18},
  number = {34},
  pages = {10595--10601},
  issn = {09476539},
  doi = {10.1002/chem.201103433},
  urldate = {2020-06-18},
  langid = {english},
}
\end{document}